\documentstyle[12pt,epsfig]{article}
\begin{document}
\title{From 3K to $10^{20}$ eV}
\author{Luis Alfredo Anchordoqui\\ 
\\
{\sl Departamento de F\'{\i}sica,}\\ 
{\sl Universidad Nacional de La Plata,}\\
{\sl C.C. 67, (1900) La Plata,}\\
{\sl Argentina}}
\date{December 18, 1998}
\maketitle

\def\thepage{\protect\raisebox{0ex}{\ } $^{{\rm {\bf PhD\,Thesis}}}_
{{\rm(summary)}}$}
\thispagestyle{headings}
\markright{\thepage}

\begin{abstract}
The problem of predicting the upper cut-off of the cosmic ray
spectrum is re-examined.
Some aspects of extremely high energy interactions and
their implications for the interpretation of giant air showers are also 
discussed.
\end{abstract}

\newpage

\pagenumbering{arabic}

\begin{flushright}
{\it To my mother}
\end{flushright}

\hfill

\section{The Why of this Thesis}

The nature and origin of
cosmic radiation has presented a challenge since its detection, some
80 years ago, in a series of pioneering ballon flights 
by Victor Hess \cite{Hess}. 
Already in 1938, it was clear from the observation of extensive air showers 
(produced by energetic cosmic rays (CRs) interacting in the high
atmosphere) \cite{PA1} 
that the CR-spectrum reached at least $10^{16}$ eV, and
continuously  running monitoring indicated that the departure from
isotropy was no greater than the statistical uncertainty of the
measurements.     
After the discovery of the microwave background 
radiation (MBR) \cite{Penzias}, it
became obvious that ultra high energy particles undergo reactions with the
relic photons (which they see as higly blue shifted), and so
extremely high energy cosmic rays (EHECRs) cannot
come from cosmologically large distances \cite{GZK}.\footnote{Usual
CR-folklore makes the following distinction: ultra-high energy CRs are
those above $10^{18}$ eV while EHECRs  refers to the ones with
energies above $10^{20}$ eV. History does not always choose the best
variables, and I shall undistinguishably refer to them,
specifing explicitly when speaking of events above $10^{20}$ eV.}
Assuming a
cosmologically homogeneus population of sources, usually referred to as 
the universal hypothesis, this interaction produces the 
Greisen-Zatsepin-Kuz'min (GZK) cut-off
around $5 \times 10^{19}$ eV. However, ingenious installations with large 
effective areas and long exposure times needed to overcome the increasingly 
low flux, have raised the
maximum observed primary particle energy to higher than $10^{20}$ eV
(see, e.g. , \cite{sd} for a recent review).
In particular, the Fly's Eye experiment recorded the world's highest
energy cosmic ray event to date, with an energy proximately to 300
EeV \cite{Bird1}. The evidence for anysotropy of EHECR arrival
directions is suggestive but statistically very weak. 
The analyses of the data sets of both Haverah Park \cite{jeremy} and
AGASA \cite{agasa}
have suggested that the events with
primary energy $> 40$ EeV reached the Earth preferentially from the general
direction of the Supergalactic Plane, a swath in the sky along which 
radio galaxies are
clustered. However, the Fly's Eye group \cite{anife} reported a small 
anisotropy towards the
Galactic Plane at energies around $10^{18}$ eV, confirmed later by the
the AGASA experiment \cite{aniaga} (two analyses that did not reveal a 
significant
correlation with the Supergalactic Plane). Recent observations dramatically 
confirms that the CR-spectrum does not end with the GZK 
cut-off \cite{takeda}.
Besides, the newest data world sample above $10^{20}$ eV has no imprint 
of possible correlations with the Galactic Plane or Supergalactic
Plane \cite{anihillas}.\footnote{A possible correlation between compact
radio quasars and the five most energetic CRs has been already 
reported \cite{qso}.}

Theoretical subtleties surrounding the production of particles with
energies below $10^{20}$ eV have
remained a puzzle.
Acceleration in astrophysical settings occurs when
charged primaries,
namely protons and heavy nuclei, achieve high energies through
repeated encounters with moving magnetized 
plasmas (``bottom up'' mechanisms) \cite{prot}.
Preferred sites are supernova shocks, galactic wind termination shocks, or
relativistic shocks associated with active galactic nuclei (AGNs) and
radio galaxies (known to be powerful sources of radio, gamma radiation,
and super-relativistic jets). Although in this last case, acceleration up
to around
$10^{21}$ eV seems to be possible by stretching the reasonable values for
the shock size and the magnetic field strength at the shock
somewhat, for the highest events, there seem to be no 
suitable extragalactic
objects such us AGNs or active galaxy clusters near the observed arrival
direction and within a
maximum distance of about 50 Mpc (the estimated angular deflection
for a proton of $10^{20}$ eV is 2.1$^\circ$). 
The gamma ray burst (GRB) phenomenon, which itself is also an oustanding
mistery in astrophysics, has been considered as one of the candidates
for the origin of EHECRs \cite{waxman}. GRBs are generally related with
catastrophic (explosive) events during which CR particles may be shocked
accelerated to extreme energies, besides, the total energy and ocurrence
rate of GRBs are in rough agreement with the production rate of EHECRs.
However, the model is not free of problems, as can be see, for instance, in
\cite{dar}.

The difficulties encountered by conventional acceleration mechanisms
in accelerating particles to the highest observed energies have motivated
suggestions in the sense that the underlying production mechanism could 
be, perhaps, of
non-aceleration nature. In the so-called ``top down'' models, charged and
neutral primaries are produced at extremely high energies, typically by
quantum mechanical decay of supermassive elementary $X$ particles related to
grand unify theories \cite{crphysrep}. These exotic particles have
also been suggested as primaries \cite{kepy}.

An alternative (perhaps mischievously) explanation for the EHECRs 
requires the breakdown of local Lorentz invariance. Namely, tiny departures 
from Lorentz invariance, too small to have been
detected otherwise, might affect elementary particle kinematics 
in such a way that some hadronic resonances which are inestable at 
low energies would become stable at very high energies. Therefore the
GZK cut-off can be relaxed or removed \cite{gm}.
It could be found some even more ``exotic'': The recently detected energetic 
particles may have shortcut their journey traversing a 
wormhole.\footnote{Naively, 
a wormhole is a tunnel in the topology of the
space-time from where in-coming causal curves can pass through and
become out-going on the other side \cite{motho}. Whether such
wormholes are actually allowed by the laws of physics is currently unknown.}
The question then could be not from where do the rays come, but from when!!!

Deepening the mistery, the particle identity is still unknown. The Fly's
Eye analysis suggested a transition from a spectrum dominated by
heavy nuclei to a predominantly light composition above a few
times $10^{19}$ eV. However, it may be worth mentioning that proton
induced air shower Monte Carlo does not fit the shower development of the
highest event with high precision. A primary heavy nucleus fits more 
closely  the
shower development and neutrinos cannot be excluded as atypical
primaries neither \cite{Hal}.
The situation is not settled down because fluctuations
in the shower development are known to be large.
On the one hand, heavy nuclei have their own merits because they can be
deflected
considerably by the galactic magnetic field which relaxes the source
direction requirements, and  they can be accelerated to higher
terminal energies because of their higher charge. On the other hand
cosmic ray nuclei with energies in excess of $2 \times 10^{20}$ eV cannot
originate from sources beyond 10 Mpc \cite{stepeleroulet,epeleroulet}. Weakly 
interacting particles such as neutrinos will have no
difficulty in propagating through the intergalactic medium, namely
the corresponding mean free path $l_\nu \approx 4 \times 10^{28}$ cm
is just above the present size of the horizon, $H_0^{-1} \approx
10^{28}$ cm \cite{nu}. Many sources of extremely high energy 
neutrinos are known, they can be produced in astrophysical objects by
the decay of pions or kaons generated as subproducts of proton-photon
interactions during the acceleration processes \cite{protheroe} or else by
``top down'' mechanisms \cite{crphysrep}. In the latter, the neutrino flux 
might extend even up to $10^{25}$ - $10^{28}$ eV. However, at these 
energies the atmosphere is  still transparent to neutrinos, and most of them
will impact on Earth \cite{ES}. Cosmic ray neutrino annihilation onto relic
neutrinos in the galactic halo was suggested as an alternative source of
EHECRs \cite{thomas}, but as the previous ones, this is not a truly
satisfactory solution \cite{waxy}. 

Summing up, the GZK cut-off provides an important constraint on the
proximity of EHECR (nucleons and nuclei) sources, thus, the origin of the 
events with energies above 100 EeV has became one of the most pressing 
questions of cosmic ray astrophysics.

This summary is ordered as follows.
The next section is
devoted to revisit the effect of MBR on EHECRs
using the continuous energy-loss (CEL). It is a matter of fact, that
even when transport equations may
be as complicated as one can afford, some simple assumptions
makes results agree, within a typical few percent, with those
genereted by numerical code, thus allowing for the isolation of
essential physics. This is exactly what is needed here in order to study
the the footprint left by the relic photons in the energy spectrum and
how can this be used to extract testeable predictions.
In Sec. 3, using
polarization measurements and synchrotron emission as a trace,
we analyze concrete candidates for testing the
acceleration models that involve hot spots and strong shock waves as
the source of the ultra-high energy component of the CR-spectrum.
Sec. 4 will deal with the influence of different hadronic models on
extensive air showers. The hadronic models considered are those
implemented in the well-known QGSJET \cite{QGSJET} and SIBYLL \cite{SIBYLL} 
event generators. The
different approaches used in both codes to model the underlying
physics are analyzed using computer simulations performed with the
program AIRES \cite{sergio}.

\section{{\it En route} to us, here on Earth}

The aim of this section is to present a simplify model of the
propagation of EHECRs which reproduces the essential elements of
the highly sophisticated computer calculations to within a few percent. 

\subsection{Energy attenuation length}

\subsubsection{\bf Nucleons}

If one assumes that the highest energy
cosmic rays are indeed nucleons, the fractional  
energy loss due to interactions
with the cosmic background radiation at temperature $T$ and redshift
$z=0$, is determined by the integral of  the nucleon loss 
energy per collision times
the probability per unit time for a nucleon collision moving 
through an isotropic gas of photons \cite{Ste68}. This integral can be
explicitly written as follows,
\begin{equation}
-\frac{1}{E} \frac{dE}{dt} =\frac{c}{2 \Gamma^2}\,\sum_j\, \int_0^{w_m} dw_r \,\, 
K_j  \, \,  
\sigma_j (w_r)\, \, w_r \, \int_{w_r/2 \Gamma}^{w_m} dw \, 
\frac{n(w)}{w^2}   
\label{conventions}
\end{equation} 
where $w_r$ is the photon energy in the rest frame of the nucleon, and
$K_j$ is the average fraction of the energy lost by
the photon to the nucleon (in the $j$th reaction channel) in the laboratory 
frame (i.e. the frame
in which the microwave background radiation is at $\approx 3$K). The
sum is carried out over all channels,
$n(w)dw$ stands for the number density
of photons with energy between $w$ and $dw$, following a Planckian
distribution \cite{COBE2} at temperature $T$, $\sigma_j(w_r)$ is 
the total cross section of the ($j$th) interaction channel, 
$\Gamma$ is the usual  
Lorentz factor of the nucleon,
and $w_m$ is the maximum energy of the photon
in the photon gas.

Thus, the fractional energy loss is given by
\begin{equation}
-\frac{1}{E} \frac{dE}{dt} = - \frac{ckT}{2 \pi^2 \Gamma^2 (c \hbar)^3}
\sum_j \int_{w_{0_j}}^{\infty}  dw_r \,
\sigma_j (w_r) \,K_j \, w_r \, \ln ( 1 -  e^{-w_r / 2 \Gamma kT})
\label{phds!}
\end{equation}
where $k$ and $\hbar$ are Boltzmann's and Planck's constants
respectively, and $w_{0_j}$ is the threshold energy for the $j$th reaction 
in the rest frame of the nucleon.

The characteristic time for the energy loss due to pair production at 
$E > 10^{19}$ eV is $t \approx 5\times 10^9$ yr \cite{BLUE}
and therefore it does not affect the spectrum of nucleons arriving from
nearby sources. Consequently, for nucleons with 
$E > 3 \times 10^{19}$eV (taking into account the interaction
with the tail of the Planck distribution), meson photoproduction is the 
dominant mechanism for energy loss. Notice that we do not 
distinguish between neutrons and protons; in addition, the inelasticity due 
to the neutron $\beta$ decay is negligible. 

In order to determine the effect of meson photoproduction on the
spectrum of cosmic rays, we first examine the kinematics of
photon-nucleon interactions. Assuming that reactions mediated 
by baryon resonances have spherically symmetric decay angular distributions 
\cite{RB}, the average energy loss of the nucleon after n resonant 
collisions is given by
\begin{equation}
K(m_{R_0}) = 1 - \frac{1}{2^n} \prod_{i=1}^{n} \left( 1 + \frac{m_{R_{_i}}^2 -
m_M^2}{m_{R_{_{i-1}}}^2} \right)
\end{equation}
where $m_{R_{_i}}$ denotes the mass of the
resonant system of the chain,  $m_M$ the mass of the associated
meson, $m_{R_{_0}} = \sqrt{s}$ is the total energy of the reaction 
in the centre
of mass, and $m_{R_{_n}}$ the mass of the nucleon.
It is well established from experiments \cite{inel} that, at very 
high energies 
($\sqrt{s}$ above $\sim 3$ GeV), the incident nucleons
lose one-half their energy via pion photoproduction independently of the
number of pions produced (leading particle effect). 

In the region dominated by baryon resonances, the cross section
is described by a sum of Breit-Wigner distributions (constructed  
from the experimental data in the Table of Particle Properties \cite{pdg})
considering the main resonances produced in
$N \gamma$ collisions with $\pi N, \pi \pi N,\, {\rm and} \, \pi K N$ 
final states. For the cross section at high energies we used 
the fits to the 
high-energy cross section $\sigma_{{\rm total}} (p\gamma)$ 
made by the CERN, DESY HERA, and COMPAS Groups \cite{pdg}.
In this energy range, the $\sigma_{{\rm total}} (n\gamma)$ is 
to a good approximation identical to $\sigma_{{\rm total}} (p\gamma)$. 

The numerical integration of Eq. (\ref{phds!}) is performed taking 
into account the
aforementioned resonance decays and the production of multipion final
states at higher centre of mass energies ($\sqrt{s} \sim 3$ GeV). 
A $\chi^2$ fit of the numerical results of equation (\ref{phds!}) 
with an exponential behaviour, $ A \, {\rm exp} \{ - B / E \}$, 
proposed 
by Berezinsky and Grigor'eva \cite{BG} for
the region of resonances, gives: $ A = ( 3.66  \pm 0.08 )
\times 10^{-8} \, {\rm yr}^{-1} $, $B =
(2.87 \pm 0.03 )\times
10^{20}$ eV with a  $\chi^2/dof = 3.9/10$. 
At high energies the 
fractional energy loss was fitted with a constant, $ C = ( 2.42 \pm 0.03 ) 
\times 10^{-8}  \,\, {\rm yr}^{-1} $ \cite{turquesa}.
These results differ from those obtained in \cite{BG}, due to a 
refined expression for the total cross section. From the values determined 
for the fractional energy loss, it is straightforward to
compute the energy 
degradation of EHECRs in terms of their flight distance.

\begin{figure}[tb]
\label{jc} 
\centering 
\leavevmode\epsfysize=8cm \epsfbox{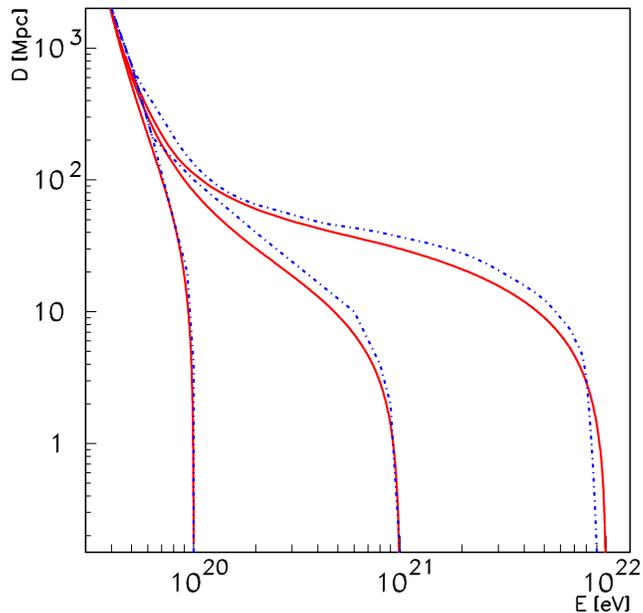}\\ 
\caption{Energy attenuation length of nucleons in the intergalactic
medium. Solid line stands for the CEL approximation, while dots for
Monte Carlo simulation which considers just single pion 
production \cite{JCronin}. }
\end{figure} 

In Fig. 1 we compare the energy attenuation length of nucleons
obtained with the CEL approximation and the Monte Carlo simulation
taken from Ref. \cite{JCronin}.
Although  the energy loss is dominated
by single pion photoproduction \cite{JCronin,Yoshida}, one can see that 
at higher energies the multiple production of pions becomes quite important.
Notice that, independently of the initial 
energy of the
nucleon, the mean energy values approach to 100 EeV after a distance
of $\approx$ 100 Mpc.
 
\subsubsection{Nuclei}

The relevant mechanisms for energy losses that extremely high energy nuclei 
suffer during their trip to the Earth are
photodisintegration and hadron photoproduction (which has a
threshold energy of $\approx 145$ MeV,
equivalent to a Lorentz factor of $10^{11}$, above
the range treated in this Thesis). 

Following the conventions of Eq. (\ref{conventions}), the 
disintegration rate of a nucleus of mass $A$ with the subsequent 
production of $i$ nucleons is given by the expression \cite{Ste69},
\begin{equation}
R_{Ai} = \frac{1}{2 \Gamma^2} \int_0^{\infty} dw \,
\frac{n(w)}{w^2} \, \int_0^{2\Gamma w} dw_r
 \, w_r \sigma_{Ai}(w_r)
\label{phdsrate}
\end{equation}
with $\sigma_{Ai}$ the cross section for the interaction. Using the 
expressions for the cross 
section fitted by Puget {\it et al.} \cite{PSB}, it is possible to work out
an
analytical solution for the nuclear disintegration rates \cite{sudafri}. 
After adding over all the possible channels for a given
number of nucleons, one obtains the effective nucleon loss rate. 
 The effective $^{56}$Fe
nucleon loss rate obtained after carrying out 
these straightforward but rather lengthy steps
can be parametrized by,
\begin{equation}
R(\Gamma)=3.25 \times 10^{-6}\, 
\Gamma^{-0.643}                                          
\exp (-2.15 \times 10^{10}/\Gamma)\,\, {\rm s}^{-1} 
\end{equation}
if $\Gamma \,\in \, [1. \times 10^{9}, 3.68 \times 10^{10}]$, and 
\begin{equation}
R(\Gamma) =1.59 \times 10^{-12} \, 
\Gamma^{-0.0698}\,\, {\rm s}^{-1}   
\end{equation}
if $ \Gamma\, 
\in\, 
[3.68 \times 10^{10}, 1. \times 10^{11}]$. 
It is noteworthy that knowledge of the iron 
effective nucleons loss rate alone is enough to obtain the corresponding 
value of $R$ for any other nuclei \cite{PSB}.

The emission of nucleons is isotropic in the
rest frame of the nucleus, and so the averaged fractional
energy loss results equal the
fractional loss in mass number of the nucleus, {\it viz.}, the Lorentz
factor is conserved. 
Because of the position of $^{56}$Fe on the binding energy curve,
it is considered to be a significant end product of stellar evolution, and
higher mass nuclei are found to be much rarer in the cosmic radiation.
Thus, hereafter, when speaking of sources of heavy nuclei we shall be thinking
on iron nuclei. The relation which determines
the attenuation length for its energy is then,  
\begin{equation}
E = E_g \,\, e^{-R(\Gamma ) \, t / 56}
\label{phdsconstraint}
\end{equation}
where $E_g$ denotes the energy with which the nuclei were 
emitted from the source, and $\Gamma = E_g / 56$.

\begin{figure}[tb]
\label{phdf3}
\centering
\leavevmode\epsfysize=7cm \epsfbox{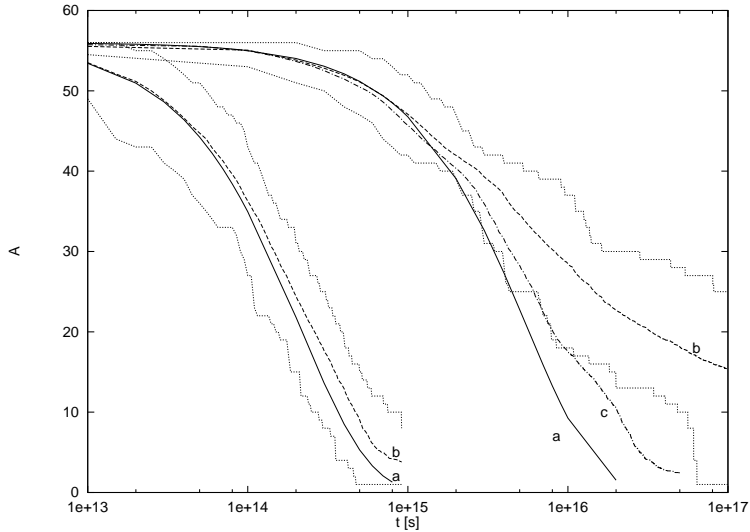}\\
\caption{Evolution of the mass number A of the heaviest fragment surviving 
photodisintegration vs. travel time for different injection energies
($\Gamma_0 = 4 \times 10^{9}$, $\Gamma_0 = 2 \times 10^{10}$). 
(a) stands for CEL
approximation. It is also included $<A>$ obtained from 
Monte Carlo simulations (with (b)
and without (c) pair creation processes) for comparison. The region
between the two dashed lines includes 95\% of the simulations. This
gives a clear idea of the range of values which can result from
fluctuations from the average behaviour.} 
\end{figure}

\begin{figure}[tb]
\label{phdf4}
\centering
\leavevmode\epsfysize=7cm \epsfbox{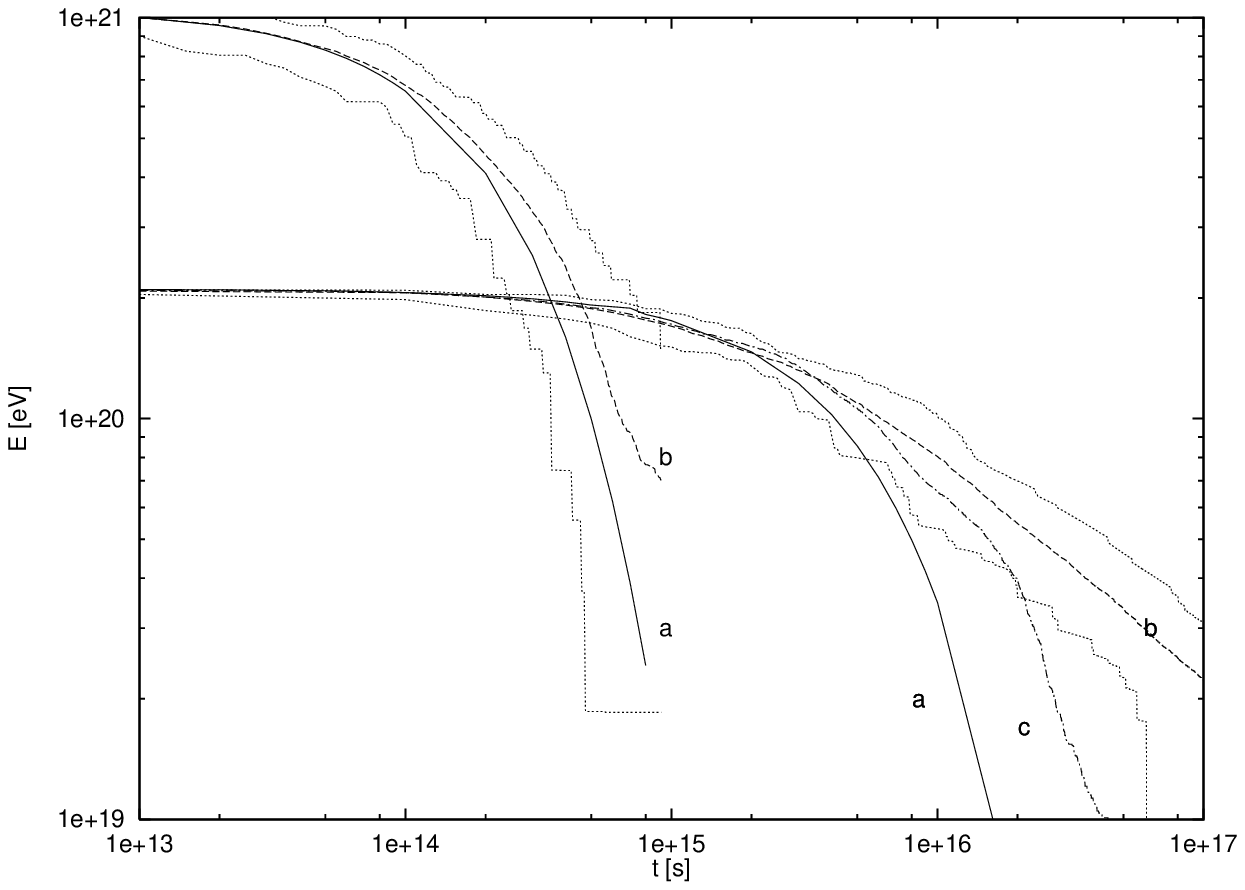}\\
\caption{Energy of the surviving nuclei ($\Gamma_0 = 4 \times
10^{9}$, $\Gamma_0 = 2 \times 10^{10}$) vs. propagation distance
obtained with the CEL (a). It is also included the energy attenuation
length computed with Monte Carlo simulations, while (b) include both 
photodisintegration and pair creation processes, (c) just the losses
due to photodisintegration. Again, the region between dashed lines
includes the 95 \% of the simulations.} 
\end{figure}

In Figs. 2 and 3 we plot the final mass (i.e. $A$) and energy $E$
of the heaviest surviving
fragment as a function of the distance for initial iron nuclei (a). For
comparison, it is also super-imposed a Monte Carlo simulation
which include the rates just discussed and also the pair creation processes
(b) \cite{esteban}. One
can see that nuclei with Lorentz factors above $10^{10}$ cannot survive 
for more than 10 Mpc (recall thet 10 Mpc correspond to $\approx
10^{15}$ s), and for these distances, the CEL approximation always lies in the
region which includes 95\% of the simulations.

At $\Gamma_0 < 5 \times 10^9$ the pair creation losses
start to be  relevant, significantly reducing the value of the Lorentz factor
 as the nucleus propagates distances of ${\cal O}$(100
Mpc). The effect has a maximum for $\Gamma_0 \approx 4 \times 10^{9}$
but becomes small again for $\Gamma_0 \leq 10^{9}$, for which
appreciable effects only appear for cosmological distances ($> 1000$
Mpc), see for instance \cite{epeleroulet}.
The efect of neglecting pair creation losses translates into keeping
$\Gamma = \Gamma_0$ constant during the propagation, and this enhances
the photodisintegration rates and then reduces $<A>$ more rapidly.
The divergences between (b) and (c) are less
pronounced in Fig. 3. Namely, for $\Gamma_0 = 4 \times 10^{9}$ the
average energies with and without pair creation proccess (obtained
with the Monte Carlo simulations) are similar up to $\approx t =
10^{16}$ s 
while the $<A>$ values sizeably differ for $t = 3 \times
10^{15}$ s onwards. When comparing with the CEL the differences arise even
earlier ($\approx$ 70 Mpc and 30 Mpc respectively). This compensation
is due to a partial
cancellation between the effects of the evolution of $\Gamma$ and of
$A$ in the values of the final energy ($E = m_p \,\Gamma \,A$), since
neglecting pair creation losses does not allow $\Gamma$ to decrease
but make instead $A$ to drop faster.

\begin{figure}[tb]
\label{slider}
\centering
\leavevmode\epsfysize=7cm \epsfbox{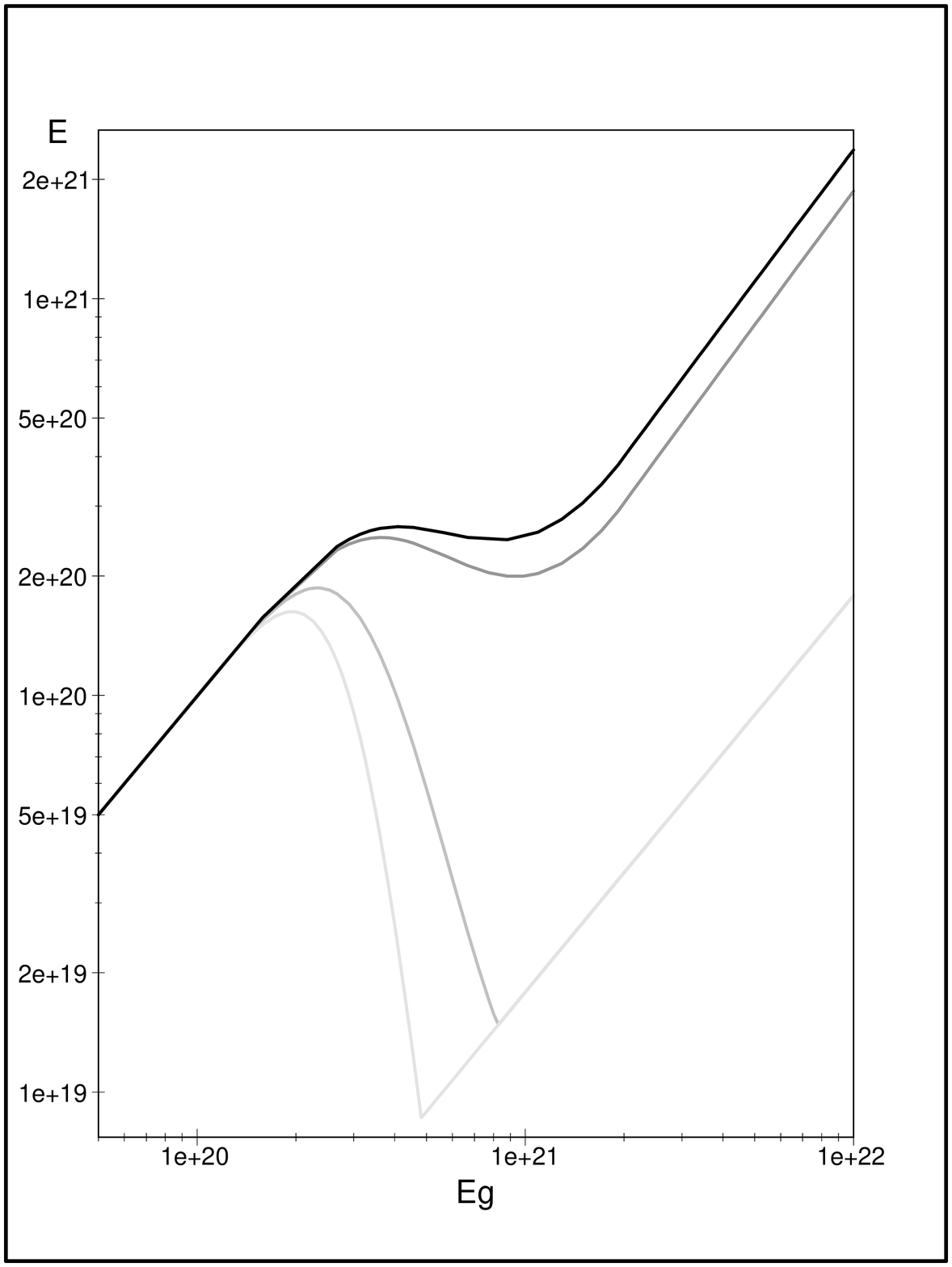}\\
\caption{Relation between the injection energy of an iron nucleus and
the final energy of the photodisintegrated nucleus for different
values of the propagation distance (from grey to black, 20 Mpc, 10
Mpc, 3.5 Mpc, and 3 Mpc.)} 
\end{figure}

In Fig. 4 we have shown the relation between the injection
energy and the energy at a time $t$ for different propagation distances. 
The graph indicates that the final energy of the nucleus is
not a monotonic function. It has a maximum at a critical energy
and then decreases to a minimum before rising again as $\Gamma$
rises, as was first pointed out by Puget {\it et al.} \cite{PSB}. 
The fact that the energy $E$ is a multivaluated function of $E_g$
leads to a pile-up in the
energy spectrum. Moreover, this
behaviour enhances a hidden feature of the energy
spectrum for sources located beyond 2.6 Mpc: A depression that preceeds 
a bump
that would make the events at the end of the spectrum (just before
the cut-off) around 50\% more probable than those in the depressed region.

\subsection{Modification of the cosmic ray spectrum}

\subsubsection{A hump follow by a cut-off}

Let us begin with the modification that the MBR produces in the 
EHECR (nucleon) spectrum. The evolution of
the spectrum is governed by the balance equation 
\begin{equation}
\frac{\partial N}{\partial t} = \frac{\partial (b(E) N)}
{\partial E} + D \,\, \nabla^2 N + Q
\label{phds@}
\end{equation}
that takes into account the conservation of the
total number of nucleons in the spectrum.
In the first term  on the right, $b(E)$ is the mean rate at which particles 
lose energy. 
The second term, the diffusion in the MBR,
is found to be extremely small due to the low density of relic
photons and the
fact that the average cosmic magnetic field is less than  $10^{-9}$ G 
\cite{mgfield} 
and is neglected in the following.
The third term corresponds to the particle injection rate into
the intergalactic medium from some hypothetical source.
Since the origin of cosmic rays is still unknown, we consider
three possible models:
1) the universal hypothesis, which assumes that cosmic rays come from
no well-defined source, but rather
are produced uniformly throughout space, 2) single point sources
of cosmic rays, and 3) sources of finite size
approximating clusters of galaxies.
In all the cases it has been assumed that nucleons propagate
in a straight line through the intergalactic medium due to the reasons 
mentioned above that allow us to neglect the diffusion term.

There exists evidence that
the source spectrum of cosmic rays has a power-law dependence
$ Q (E) = \kappa \, E^{- \gamma}$ (see for instance \cite{Hill}). 
With the hypothesis that cosmic rays are produced from sources located
uniformly in space, with this power law energy dependence (which
implies a steady state process), the solution of Eq. (\ref{phds@}) is
found to be 
\begin{equation} 
N(E) = \frac{Q(E) \, E }{ b(E)\, (\gamma - 1)}.
\end{equation} 

For the case of a single point source,
the solution 
of equation (\ref{phds@}) reads \cite{Stk},
\begin{equation}
N(E, t) = \frac{1}{b(E)} \, \int_E^{\infty} \, Q(E_g, t') \,dE_g, 
\label{phdsspectrum}
\end{equation}
with
\begin{equation}
t'=  t - \int_E^{E_g} \frac{d\tilde{E}}{b(\tilde{E})},
\label{phdsespectro}
\end{equation}
and $E_g$ the energy of the nucleon when emitted by the source.
The injection spectrum of a single  source located at $t_0$ from the observer
can be written as $Q(E, t) \,= \,\kappa \, E^{- \gamma} \,
\delta(t - t_0)$ and for simplicity, we consider the distance as measured
from the source, that means $t_0 = 0 $. 
At very
high energies, i.e. where $b(E) = C \, E$ (with $C$  the constant
defined above in the discussion of fractional energy loss) 
the total number of particles at a given distance from the source is 
given by
\begin{equation}
N(E, t) \approx \frac{\kappa}{b(E)} \int_E^{\infty}\, E_g^{-\gamma} \,
\, \, \delta\left( t - \frac{1}{C} \ln 
\frac{E_g}{E} \right) dE_g,
\end{equation}
or equivalently,
\begin{equation}
N(E, t) \approx \kappa \, E^{-\gamma} e^{- \, (\gamma -1) \, C \, t }.
\end{equation}
This means that the spectrum is uniformly damped by a factor depending
on the proximity of the source.

At low energies, in the region dominated by baryon resonances, the
parametrization of $b(E)$ does not allow a complete analytical solution.
However using the change of variables, 
\begin{equation}
{\tilde t} =
\int_E^{E_g} \frac{d\tilde{E}}{ b(\tilde{E})},
\end{equation}
with $E_g = \xi (E, \tilde{t})$ and $d{\tilde t} =
dE_g/b(E_g)$, we easily obtain,
\begin{equation}
N(E, t) =  \frac{\kappa}{b(E)} \int_0^{\infty}\, \xi(E, \tilde{t})^
{-\gamma}\,
\,\delta({\tilde t} - t) \,\, b [ \xi(E, {\tilde t}) ] \,\,d{\tilde t},
\end{equation}
and then, the compact form,
\begin{equation}
N(E, t) = \frac{\kappa}{b(E)} E_g^{-\gamma} b(E_g).
\label{hojo}
\end{equation}
In our case, where we have assumed an exponential behaviour of 
the fractional loss energy, $E_g$ is fixed by the constraint:  
$\,
A \, t \, - \,  {\rm Ei}\,(B/E) 
+ \, {\rm Ei}\, (B/E_g) 
= 0
$,  Ei being the exponential integral \cite{Abra},
and $B$ the constant defined above in the parametrization of
Berezinsky and Grigor'eva.

Studies are underway of the case of a nearby source 
(i.e. within about 100 Mpc)
for which 
the fractional energy loss $\Delta E/E$ is small. Writing $E_g =
E + \Delta E$, and neglecting higher orders 
in $\Delta E$,
it is possible to obtain an analytical solution
and the most relevant characteristics of the modified spectrum.
Therefore, an expansion of
$b(E + \Delta E)$ in powers of $\Delta E$ in Eq. (\ref{phdsespectro}) 
allows one to obtain,
an expression for
$\Delta E$,
\begin{equation}
\Delta E \approx \left\{ {\rm exp}\left[ \frac{t \, b(E)\,
(E + B)}{E^2} \right] - 1 \right\}
\frac{E^2}{(E + B)}. 
\label{phdsl1}
\end{equation}

To describe the modification of the spectrum, 
it is convenient to 
introduce the factor $\eta$, the ratio between the
modified spectrum and the unmodified one, that results in 
\begin{equation}
\eta = \left( \frac{E + \Delta E}{E}\right)^{-\gamma} 
\frac{b(E+\Delta E)}{b(E)}.
\label{phdsl2}
\end{equation}  
Equations (\ref{phdsl1}) and (\ref{phdsl2}) describe the spectral modification
factor up to energies of $\approx 95$ EeV ($\approx
85$ EeV) with a precision of
4\% (9\%) for a source situated at 50 Mpc (100 Mpc).

\begin{figure}[tb]
\label{eta} 
\centering 
\leavevmode\epsfysize=9cm \epsfbox{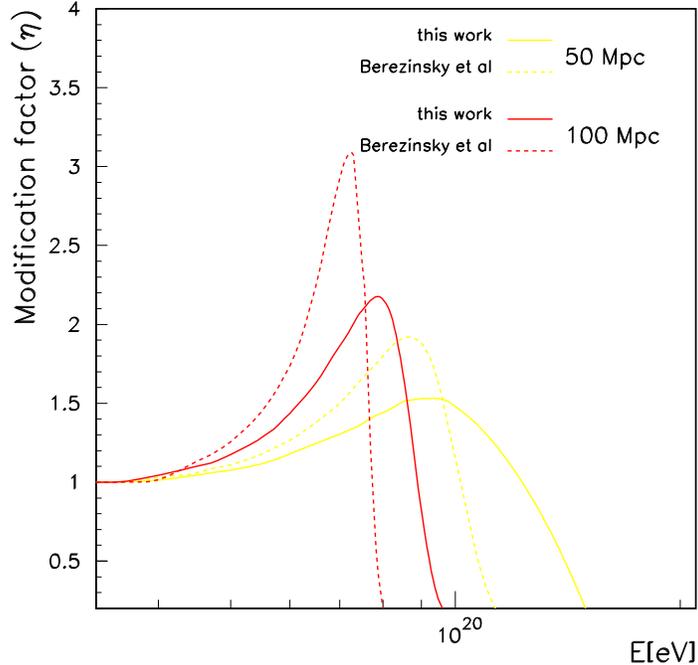}\\ 
\caption{Modification factor of single-source energy spectra for
different values of propagation distance and power law index
$\gamma = 2.5$.}
\end{figure}

In Fig. 5 we plot the modification factors for the case of 
near sources with power law injection ($\gamma = 2.5$) to 
compare with the corresponding results of \cite{BG}. The most 
significant features of $\eta$ are the 
bump and the cut-off. It has been
noted in the literature that the continuous energy loss approximation
tends to overestimate the size of the bump \cite{A,Lee}. Fig. 5 shows 
how the bump is less pronounced in our treatment. 
As the bump is a consequence of the
sharp (exponential) dependence of the free path of the nucleon on energy, 
we attribute these differences to the
replacement of a cross section 
approximated by
the values at threshold energy \cite{BG}, by a more detailed expression
taking into account the main baryon resonances that turn out to be important.

Another alternative for the particle injection
comes from clusters of galaxies. These are usually modelled by a set of point sources with 
spatially uniform distribution, although in reality the 
distribution of galaxies inside the clusters is somewhat non-uniform. 
In our treatment we shall assume that the concentration of potential
sources at the center of
the cluster is higher than that 
in the periphery, and we adopt a spatial 
gaussian distribution.
With this hypothesis, the particle injection rate into the
intergalactic medium is given by  
\begin{equation}
Q(E,t) = \kappa \int_{-\infty}^{\infty} \frac{E^{-\gamma}}{\sqrt{2 \, \pi}\, 
\sigma} \, \delta(t-T) \; 
{\rm exp}\left\{ \frac{-(T-t_0)^2}{2\,\sigma^2} \right\} dT
\end{equation}
A delta function expansion around  $t_0$, with derivatives denoted
by lower case Roman superscripts,
\begin{equation}
\delta(t-T) = \delta(t-t_0) + \delta^{(i)}(t-t_0) (T-t_0)+\frac{1}{2!}
\delta^{(ii)}(t-t_0) (T-t_0)^2 + \dots
\end{equation}
leads to a convenient form for the injection spectrum, which is given
by,
\begin{equation}
Q(E,t^\prime) =  \kappa E^{-\gamma} \,[\,\delta(t^{\prime}-t_0) + 
\frac{\sigma^2}{2!} \, 
\delta^{(ii)}(t^{\prime}-t_0) + \frac{\sigma^4}{4!}\, \delta^{(iv)}(t^{\prime}-t_0)
+\dots]
\end{equation}
From Eqs. (\ref{phdsspectrum}) and (\ref{phdsespectro}), it is
straightforward to compute an expression for the modification factor,
\begin{equation}
\eta = \frac{E_g^{-\gamma}\, b(E_g)}{E^{-\gamma}\,b(E)} \, 
 \left\{ 1+\frac{ \sigma^2 A^2 e^{-2B/E_g}}{2!} F_1(E_g) + \frac{\sigma^4 A^4 e^{-4B/E_g}}
{4!} F_2(E_g)  + {\cal O}(6) \right\},
\label{hoj}
\end{equation}
where
\begin{displaymath}
F_1(E_g)= 2 B^2 E_g^{-2} + 
(2-3\gamma) B E_g^{-1} +(1-\gamma)^2,
\end{displaymath}
and
\begin{eqnarray}
F_2(E_g) & = & 24 B^4 E_g^{-4} + (4 - 50
\gamma) B^3 E_g^{-3} + (35 \gamma^2 -25 \gamma +8) B^2 E_g^{-2}
\nonumber \\
& + & (-10
\gamma^3 +20 \gamma^2 - 15 \gamma +4) B E_g^{-1} + (1- \gamma)^4. \nonumber
\end{eqnarray}

\begin{figure}[tb]
\label{virgo} 
\centering 
\leavevmode\epsfysize=8cm \epsfbox{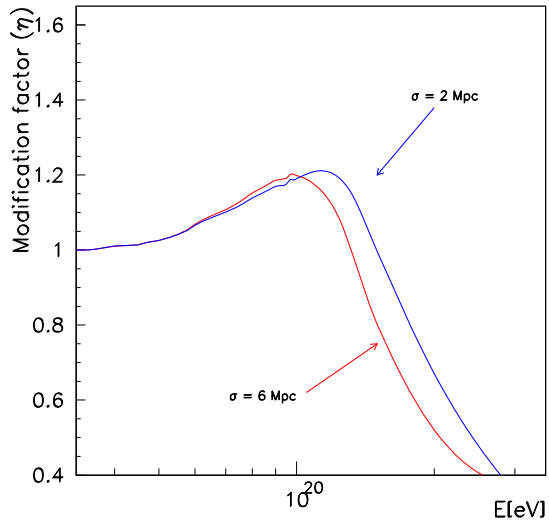}\\ 
\caption{Modification factor of extended-source energy spectrum for a
propagation distance $\approx$ 18.3 Mpc and power law index $\gamma$
= 2.5.}
\end{figure}

The modification factor for the case of extended sources
described by Gaussian distributions of widths
2 and 6 Mpc
at a distance of 18.3 Mpc is shown in Fig. 6. 
These may be taken as very crude models of the Virgo cluster \cite{Virgo},
assuming that there is no other significant energy loss mechanism
for cosmic rays traversing parts of the cluster beyond those due
to interactions with the cosmic background radiation.

A power law injection ($\gamma = 2.5$) was used again, as in the pointlike
case.
The curves can be understood qualitatively. Both peaks are in about
the same place, occurring around the onset of pion photoproduction.
The broader source distribution
reflects the losses suffered by the more remote part of the distribution
in traversing a greater distance to us. 

The high energy cosmic ray spectrum as measured by Fly's Eye
detector, both in stereo and monocular modes, has been fitted 
to Eq. (\ref{hojo}) and  $\kappa \,E^{-\gamma} \,\times$
Eq. (\ref{hoj}). The results are shown in Table I.  
The first two rows do not include the experimental energy 
resolution \cite{hojvat} while the third one
does \cite{tere}.\footnote{It is important to stress that the if the most energetic event is
included, the single source as well as the extended source model,
cannot explain the data.}

\begin{table}
\begin{center}
\caption{Results of the fits performed by Dova {\it et al.} [53,54].}
\hfill
\begin{tabular}{ccccc}
\hline
\hline
source & $\gamma$ & log $\kappa$ & distance [Mpc] & $\sigma$ [Mpc]\\
\hline
single  & 3.27 $\pm$ 0.03 & 29.64 $\pm$ 0.02 &
$113^{+10}_{-20}$ & $\dots$ \\
extended  & 3.27 $\pm$ 0.03 & 29.64 $\pm$ 0.02 & 93 $\pm$ 15 &
30 $\pm$ 10 \\ 
extended & 3.27 $\pm$ 0.02 & 29.30 $\pm$ 0.45 & 119 $\pm$ 33 & 35 \\
\hline
\end{tabular}
\end{center}
\end{table}

\subsubsection{\bf A depression before a bump on the EHECR nuclei spectrum}

The photodisintegration process
results in the production of nucleons of ultrahigh energies
with the same Lorentz factor of the parent nucleus. 
As a consequence, the total number of particles is
not conserved during propagation. However, the solution of the problem
becomes quite simple if we separately
treat both the evolution of the heaviest fragment and those fragments 
corresponding 
to nucleons emitted from the travelling nuclei.
The evolution of the differential spectrum of the 
surviving fragments is again governed by a
balance equation 
that takes into account the conservation of the
total number of particles in the spectrum.
Using the formalism presented in the previous section, 
and considering the case of a single  source located at $t_0$ 
from the observer, with 
injection spectrum $Q(E_g, t) \,= \,\kappa \, E_g^{- \gamma} \,
\delta(t - t_0)$,  
the number of particles with energy $E$ at time $t$ is given by,
\begin{equation}
N(E, t) dE  = \frac{\kappa E_g^{-\gamma+1}}{E} dE, 
\label{phdsespectros}
\end{equation}
with $E_g$ fixed by the constraint (\ref{phdsconstraint}).

Let us now consider the evolution of nucleons generated by 
decays of nuclei during their propagation. 
For Lorentz factors less than  $10^{11}$
and distances less than 100 Mpc the energy with which the secondary
nucleons are produced is approximately equal to the energy with which they
are detected here on Earth. The number of nucleons with energy $E$ at time
$t$ can be approximated by the product of the number of nucleons generated
per nucleus and the number of nuclei emitted.
When the nucleons are emitted with energies above 100 EeV
the losses by meson photoproduction start to become significant.
However, these nucleons come from heavy nuclei with 
Lorentz factors $\Gamma  > 10^{11}$ which are completely disintegrated
in distances  less than 10 Mpc. Given that the mean free path of the 
nucleons is about $\lambda_n \approx 10$ Mpc, it is reasonable to 
define a characteristic time   $\tau_{_{\Gamma}}$ given by the
moment in which the number of nucleons is reduced to $1/e$ of its 
initial value $A_0$. In order to determine the modifications of the spectrum 
due to the losses which the nucleons suffer due to interactions with the relic
photons, we assume that the iron nucleus emitted at $t = t_0$ 
is a travelling source which at the end of a time  $\tau_{_{\Gamma}}$
has emitted the 56 nucleons altogether. In this way the injection 
spectrum of nucleons ($\Gamma \approx 10^{11}$) can be approximated by,
\begin{equation}
q(E_G,t) =  
\kappa \, A_0^{-\gamma+1} \, E_G^{-\gamma} \delta(t - \tau_{_{\Gamma}}),  
\end{equation}
where $A_0$ is the mass of the initial nucleus and the energy with
which the nucleons is generated is given by
$E_G = E_g / A_0$.
 
The number of nucleons with energy $E$ at time $t$ is 
given by,
\begin{equation}
n(E,t) dE = \frac{\kappa \, A_0^{-\gamma+2} \, E_G^{-\gamma+1}}{E} dE   
\end{equation}
and the relation between injection energy and the energy at time $t$ remains
fixed by the relation, $\, A \, (t - \tau_{_{\Gamma}}) \, - \,  
{\rm{Ei}}\,(B/E) 
+ \, {\rm{Ei}}\, (B/E_G) 
= 0
$,  Ei being the exponential integral,
and $A$, $B$ the parameters of the fractional energy loss of 
nucleons fitted in the previous section. 
 
\begin{figure}[tb]
\label{3etafe} 
\centering 
\leavevmode\epsfysize=8cm \epsfbox{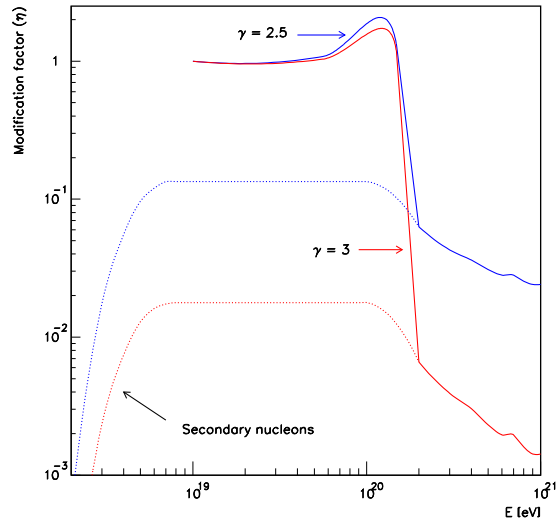}\\ 
\caption{Modification factors of heavy nuclei sources located at 20
Mpc.}
\end{figure}

In Fig. 7 it is shown the
modification factors for the case of sources of iron nuclei (propagation
distance 20 Mpc) together with the spectra of secondary nucleons. It is
clear that the spectrum of secondary nucleons around the pile-up is
at least one order of magnitude less than the one of the surviving
fragments.
In Figure 8 we have 
plotted the modification factor
for a source of nuclei located at 2.8 Mpc. It displays a bump and a 
cut-off and, in addition, 
a depression before the bump. 
A detailed analysis of the variation of these features with the
propagation distance and the spectral index was presented in \cite{prd2}.
It is important to stress that the
mechanism that produces the pile-up, which can be seen
in Figs. 7 and 8, is completely different to the one that produces
the bump in the case of nucleons (Figs. 5 and 6).

\begin{figure}
\label{phdnu}
\centering
\leavevmode\epsfysize=7cm \epsfbox{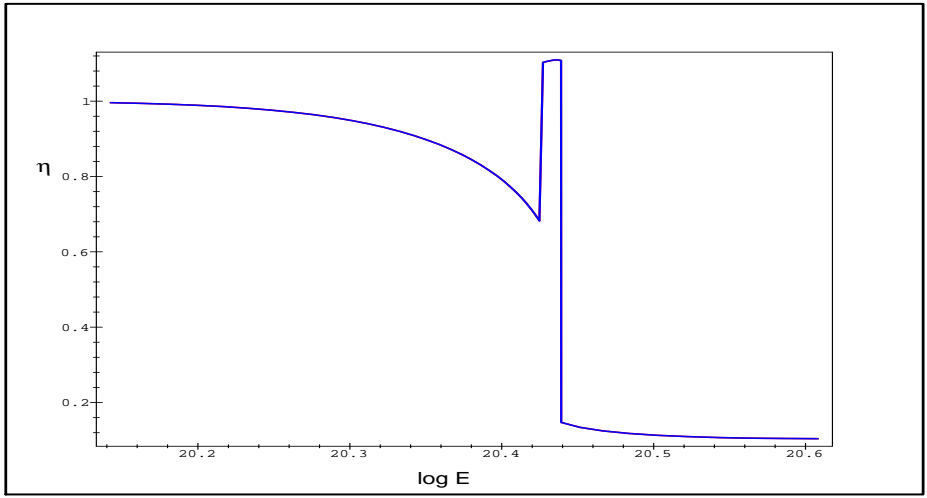}\\
\caption{Modification factor of single source energy spectrum for a
propagation distance of 2.8 Mpc, assuming a differential power law
injection spectrum with spectral index $\gamma$ = 2.5.} 
\end{figure}

In this last case, the photomeson production involves
the creation of new particles that carry off energy, yielding
nucleons with energies ever closer to the photomeson production threshold. 
This mechanism, modulated by the fractional energy loss, is responsible for
the bump in the spectrum.
The cut-off is a consequence of the conservation of the number
of particles together with the properties of the injection spectrum 
($\int_{E_{_{\rm th}}}^\infty 
E_g^{-\gamma} dE_g< \infty$).

In the case of nuclei, since the
Lorentz factor is conserved, the surviving fragments see the photons of 
the thermal
background always at the same energy.
Then, despite the fact that nuclei injected with energies over the 
photodisintegration threshold lose energy by losing mass, they never 
reach the threshold.
The observed pile-up in the modification factors is due solely to the 
multivalued nature of the energy
at time $t$ as a function of the injection energy: Nuclei injected with 
different energies can arrive with the same energy but with different masses.

It is clear that, except in the region of the pile-up, the modification
factor $\eta$ is less than unity, since $\eta = (E/E_g)^{\gamma-1}$. 
This assertion seems to be in contradiction with the conservation of 
particle number. 
Actually, the conservation of the Lorentz factor implies,
\begin{equation}
\kappa \,E_g^{-\gamma}\,dE_g|_{_\Gamma} = N(E,t)\, dE|_{_\Gamma}
\label{phdses}
\end{equation}
in accord with the conservation of the number of particles in
the spectrum. Moreover, the condition (\ref{phdses}) completely determines 
the evolution of the energy spectrum of the surving fragments 
(\ref{phdsespectros}). 
Note that in order to compare the modified and unmodified spectra, with 
regard to  conservation of particle number, one has 
to take into account that the corresponding energies are shifted. As follows 
from (\ref{phdses}), the
conservation of the number of particles in the spectrum is
given by,
\begin{equation}
\int_{E_{_{\rm th}}}^{E_{\pi_{\rm th}}} N(E,t) dE = 
\int_{E_{g_{\rm th}}}^{E_{g{\pi_{\rm th}}}} \kappa\,E_g^{-\gamma}\ dE_g
\end {equation}
with $E_{_{\rm th}}$ and $E_{\pi_{\rm th}}$ the threshold energies
for photodisintegration, and photopion production processes respectively.

Let us now return to the analysis of Fig. 4
in relation to the depression in the spectrum.
In the case of a nearby iron source, located around  3 Mpc, and 
for injection energies below the multivalued region of the 
function $E (E_g)$, $E$ is clearly less than $E_g$ 
and, as a consequence the depression in the 
modification factor is apparent.  
Then, despite the violence of the photodisintegration process via the 
giant dipole resonance, for nearby sources none of the 
injected nuclei are completely  
disintegrated yielding this unusual depression before the bump.
For a flight distance of 3 Mpc, the composition of the arrival nuclei
changes from $A=50$ (for $\Gamma \approx 10^9$) to $A=13$ (for
$\Gamma \approx 10^{11}$).
However, the most important variation takes place in the region of the bump,
where $A$ runs from 48 to 13, being heavy nuclei of $A=33$ the most abundant.
For propagation distances greater than 10 Mpc one would expect just 
nucleons to arrive for injection energies above $9 \times 10^{20}$ eV. In this 
case the function becomes multivalued below the
photodisintegration threshold and then there is no depression at all.
For an iron source located at 3.5 Mpc, the
depression in the spectrum is almost invisible $({\cal O}(1\%))$, 
in good agreement with the
results previously obtained by Elbert and Sommers using  Monte Carlo
simulation \cite{ES}.

The structure of the spectrum, recently published by AGASA
experiment (Fig. 9) \cite{takeda}, seems to suggest the presence of a bump 
around $4 \times 10^{19}$ eV; the central points may be unvailing a second 
bump around $2 \times
10^{20}$ eV. Whatever the source of the highest energy CRs, because
of their interactions with the 2.7K relic radiation and magnetic
fields in the universe, those reaching the Earth will have spectra,
composition and arrival directions affected by propagation.
Within the universal hypothesis, nucleons with energies above 100 EeV are pushed
below this limit, and as a consequence, particles which
originally had a higher energy pile-up forming a bump around $5 \times
10^{19}$ eV followed by a sharped cut-off. 
While the origin of the highest energy cosmic rays is
still uncertain, it is not necessary to invoke top down 
mechanisms to
explain the existing data. It seems likely that shock
acceleration at Fanaroff-Riley radio galaxies, and Starburst
galaxies located around 3 Mpc away (recall that nucleus energy, for
$\Gamma > 10^{11}$ is degraded even faster than 
that of protons) can account for the existing data at the end of 
the CR-spectrum (energies above 100 EeV).
The existing experimental data are insufficient to draw
any definitive conclusions, and we are obliged to
present this idea as a hypothesis to be tested by experiment.

\begin{figure}[tb]
\label{taky} 
\centering 
\leavevmode\epsfysize=8cm \epsfbox{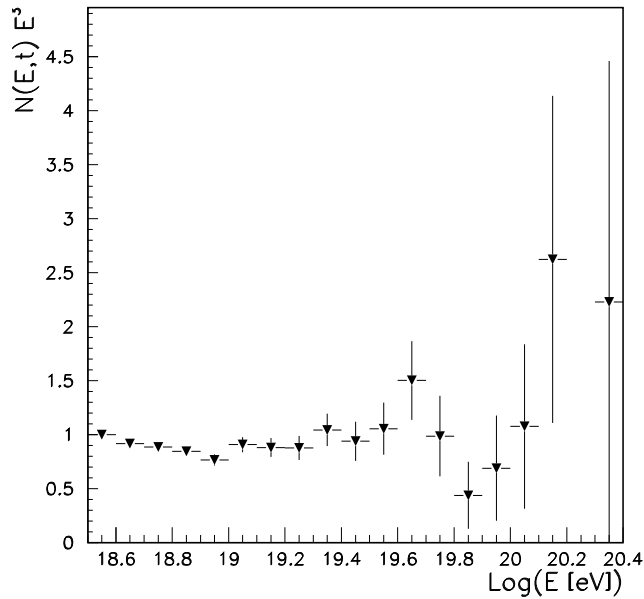}\\ 
\caption{Energy spectrum of EHECRs measured by AGASA experiment.
The spectrum is normalized to the first event ($E = 10^{18.55}$ eV).}
\end{figure}

\section{The southern CR-sky}

Radio galaxies considered as sources of CR  beyond the 
        GZK cut-off 
must be quite close to our own galaxy ($z \leq 0.03$). There are just a few
objects within this range and, consequently (if these are
responsable of the end of the CR-spectrum), the expected EHECR
distribution on
the sky should be highly anisotropic. Assuming that the intergalactic
magnetic field is near 1 nG, it should be expected an excess of CR
detections at energies larger than $5\times10^{19}$ eV within regions of
angular radius $\theta\leq 15^{\circ}$ and centered at the positions of
the nearest active galaxies.
According to this picture, the southern
CR-sky should be dominated by three outstanding sources: Centaurus A (the
nearest radio galaxy) which may provide the most energetic particles
detectable on Earth, Pictor A (a strong source with a flat
radio spectrum) which would contribute with the larger CR flux
\cite{RB}, and PKS 1333-33 which might be a source of events similar to
those recently detected in the Northern Hemisphere. In addition to these
sources, there are other two southern candidates, Fornax A ($z=0.057$)
and PKS 2152-69 ($z=0.027$), which
could provide contributions to the CR flux above the
cut-off.\footnote{Concerning nuclei sources, it should be mentioned
NGC 253.} An approximate
theoretical picture
of the up to now unexplored ultra high energy CR southern sky is
consequently at our disposal. In this section we shall discuss the main
features of the radio-galaxies Cen A and PKS 1333-33. 
Before doing so, let us
briefly summarize diffuse shock acceleration mechanisms.

\subsection{Bottom up models}

The hot spots of extended radio sources are regions of strong synchrotron
emission \cite{mei}. These regions are produced when the bulk
kinetic energy of the jets ejected by a central active source (supermassive
black hole + accretion disk) is reconverted into relativistic particles
and turbulent fields at a ``working surface'' in the head of the jets
\cite{blare}. The speed $v_{\rm h}$ with which the head of a jet advances
into the intergalactic medium of particle density $n_{\rm e}$ can be obtained 
by balancing the momentum flux in the jet against the momentum flux of the 
surrounding medium. Measured in the frame comoving with the advancing head,
$v_{\rm h}\approx \;v_{\rm j}\,[ 1 + ( n_{\rm e} /n_{\rm j})^{1/2}]^{-1}$,
where $n_{\rm j}$ and $v_{\rm j}$ are the particle density and the velocity
of the jet flow, respectively. Clearly, $v_{\rm j}> v_{\rm h}$ for 
$n_{\rm e} \geq n_{\rm j}$, in such a way that the jet will decelerate. The 
result is the formation of a strong collisionless shock, which is responsible 
for particle reacceleration and magnetic field amplification. 
The acceleration of particles up to ultrarelativistic energies in the hot
spots is the result of repeated scattering back and forth across the shock
front. The particle deflection in this mechanism is produced by Alfv\'en waves 
in the turbulent magnetic field. 
This process has been studied in detail by
Biermann and Strittmatter \cite{BS}. Assuming that the energy density
per unit of wave number of MHD turbulence is of Kolmogorov type, i.e. 
$I(k) \; \alpha \; k^{-s}$ with $s= 5/3$, the acceleration time
scale for protons given by:
\begin{equation}
E \left( \frac{dE}{dt} \right)_{_{\rm ACC}}^{-1} \; \approx \; 
\frac{40}{\pi
\, c} \;\; \beta_{_{\rm JET}}\,\!\!\!\!\!\!\!^{-2} \;\; u \;\; 
R_{_{\rm HS}}\,\!\!\!\!\!^{2/3} \, \left(
\frac{E}{e \, B} \right)
\end{equation}
where $\beta_{_{\rm JET}}$ is the jet velocity in units of $c$, $u$ is
the ratio of turbulent to ambient magnetic energy density in the hot
spot (of radius $R_{_{\rm HS}}$), and $B$ is the total magnetic field
strength. The acceleration process will be efficient as long as the
energy losses by synchrotron radiation and photon--proton
interactions do not become dominant. Considering an average cross
section $\bar{\sigma}_{\gamma p}$ for the three dominant
pion--producing interactions \cite{armstrong}, 
$\gamma + p \rightarrow p + \pi^0$, 
$\gamma + p \rightarrow n + \pi^+$,
$\gamma + p\rightarrow p + \pi^+ + \pi^- $, we get the time scale of
the energy losses within a certainty of 80\%:
\begin{equation}
E  \left( \frac{dE}{dt} \right)_{_{\rm LOSS}}^{-1} \; \approx \;
\frac{6 \, \pi \, m_p^4 \, c^3}{\sigma_{_{\rm T}} \, m_e^2 \, B^2 \, 
(\,1 \,+\,
a\, A\,)}\; E^{-1}
\end{equation} 
where $a$ stands for the ratio of photon to magnetic energy densities, 
$ \sigma_{_T}$ is the classical Thomson cross section, 
and $A$ gives a measure of the relative strength of $\gamma p $
interactions  against the synchrotron emission. Biermann and
Strittmatter \cite{BS} have estimated $A \approx 200$, almost
independently of the source parameters. The most energetic protons
injected in the intergalactic medium will have an energy that can be
obtained by balancing the energy gains and losses:
\begin{equation}
E_{{\rm max}}  =  7.8\times 10^5 \;\beta_{_{\rm jet}}\;
\!\!\!\!\!\!\!^{\;3/2}
\;u^{3/4}\;R_{_{\rm hs}}^{\;\;-1/2} \,\, B_{-5}^{-5/4}
\; ( 1 + A\,a)^{-3/4} \;\,{\rm EeV},
\label{phdsemaxim}
\end{equation}
here $B_{-5}$ is the magnetic field in units of $10^{-5}$ G. 

The diffusive shock acceleration 
process in the working surface leads to a power law particle 
spectrum \cite{bell},
\begin{equation}
Q(E)= \kappa \; E^{-\gamma}\;\;  (E_0< E <E_{{\rm max}})
\label{qespec}
\end{equation}
where $\kappa =(n_0/\gamma -1)\;E_{0}^{1-\gamma}$ for $E_{\rm max}\gg
E_{0}$,  
and $n_0$ is the particle density in the source.

\subsection{Centaurus A}

The most attractive features of Centaurus A (a complex and extremely 
powerful 
radio source identified at optical frequencies with the galaxy NGC
5128) as a possible source of CRs are its active nature, its
large, non-thermal radio lobes, and its proximity $\sim$ 
3.5 Mpc.\footnote{Centaurus A (Cen A) was suggested as a
possible source of EHECRs by Cavallo \cite{cavallo}, 
from quite general energetic arguments.}  
Radio
observations at different wavelengths \cite{14,Junkes} show a structure
composed by a compact core, a one--sided jet, Double Inner Lobes, a
Northern Middle Lobe, and two Giant Outer Lobes (See Fig. 10). 
This morphology,
together with the polarization data obtained by Junkes {\em et al.}
\cite{Junkes} and the large--scale radio spectral index distribution 
\cite{RC}, 
strongly support the picture
of an active radio galaxy with a jet forming a relatively small angle
($\approx 30^\circ$)
with the line of sight. The jet would be responsible for the
formation of the Northern Inner and Middle Lobes when interacting with
the interstellar and intergalactic medium, respectively. The Northern
Middle Lobe can be interpreted as a ``working surface'' \cite{blare} at
the end of the jet, a place where strong shocks are produced by
plasma collisions, i.e. it can be considered as the hot spot of a
galaxy with a peculiar orientation.

\begin{figure}[tb]
\label{jorge} 
\centering 
\leavevmode\epsfysize=11cm \epsfbox{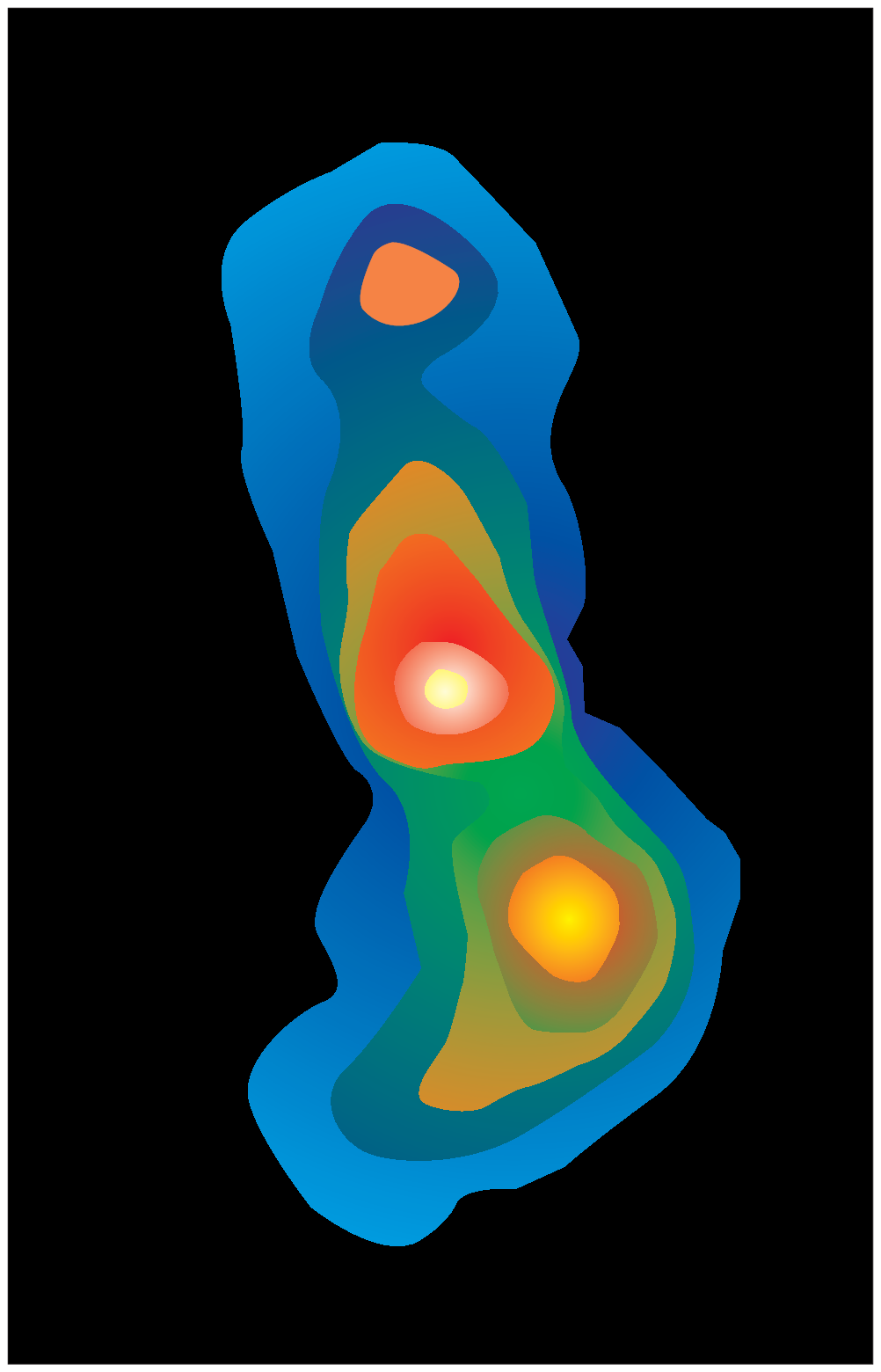}\\ 
\caption{Continuous radio emission from Cen A (1420 MHz). Brightness
temperature runs from .1K to 50K. The background has
been substracted (data taken from \cite{haslam}).}
\end{figure}

Subluminal velocities of $\approx 0.15 c$ have been detected in the
jet \cite{tingay}. 
For $\theta \approx 30^\circ$, we have $\beta_{_{\rm JET}}\approx 0.3$.
One can estimate $u \approx 0.4$ from the radio 
spectral
index of the synchrotron emission in the Northern Middle Lobe and the
observed degree of linear polarization in the same region.  
The size of the hot spot can be directly measured
from the large--scale map obtained by Junkes {\em et al.} \cite{Junkes}
with the assumed distance of 3.5 Mpc, giving as a result $R_{_{\rm HS}}
\approx  1.75$ kpc. A
value $a\sim0.01$ seems to be reasonable for a source with the 
luminosity of Cen A \cite{BS}.

The equipartition magnetic field can be obtained for the pre-shock region
from the detailed radio observations by Burns {\em et al.} \cite{14}. The field 
component parallel to the shock will be amplified in the post-shock region by
a compression factor $\xi$. In the case of a strong, nonrelativistic shock 
front, $\xi\rightarrow 4$ \cite{lev}, and then, if 
$B_{\parallel}\sim B_{\perp}$, we have,
$B  \approx (\xi ^{1/2} + 1 )^{1/2}\;B_{\perp} \approx  5\times 10^{-5}\;\;\;{\rm G}$.
An enhancement of the $B_{\parallel}$ component in the Northern Middle Lobe
can be clearly seen in the polarization maps displayed in Ref. \cite{Junkes}.

 With the above mentioned values in favor of the input parameters of 
Eq. (\ref{phdsemaxim}), the maximum energy of the protons injected in the 
intergalactic space results $E_{{\rm max}} \sim 10^{21}$ eV \cite{cena}.

We can infer the index in the power-law spectrum from multifrequency 
observations of the synchroton radiation produced by the leptonic component of 
the particles accelerated in the source (see the standard
formulae, for instance, in the book by Pacholczyk \cite{22}). 
Using the radio spectral index obtained by Combi and Romero
for the hot spot region \cite{RC}, we get $\gamma=2.2$.

Spectral modifications that arise from the interaction of extremely
high energy protons with the MBR can be computed using Eqs. (\ref{phdsl1})
and (\ref{phdsl2}) with
sufficient accuracy in the case of a nearby source like Cen A. Fig. 11
shows the modification factor $\eta$. It can be appreciated that the
spectrum is not dramatically modified, as in the case of most distant
sources.

Finally, it should be mentioned that some observational reports seem
to back up the above-outlined model. $\gamma$-rays in the energy range
300 - 3000 GeV from Cen A have been detected by Grindlay {\it et al.}
in 1975 with the Narrabi optical intensity interferometer
\cite{grindlay}. An excess of cosmic radiation from the direction of
Cen A in the range $10^{15}$ - $10^{17}$ eV has been reported by Clay
{\it et al.} \cite{clay}.

\begin{figure}[tb]
\label{screta} 
\centering 
\leavevmode\epsfysize=9cm \epsfbox{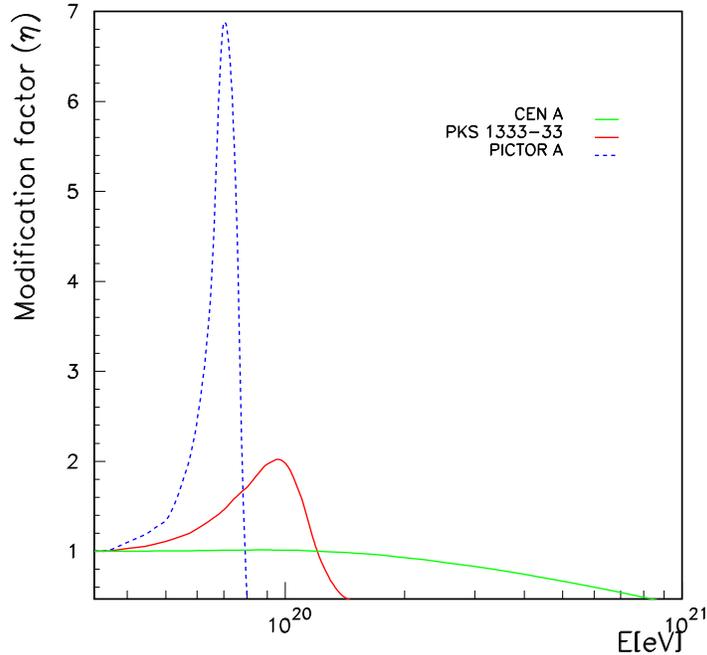}\\ 
\caption{Modification factors of southern radio-galaxies. The parameters
of Pictor A were taken from Ref. \cite{RB}.}
\end{figure}

\subsection{PKS 1333-33}

The radio galaxy PKS 1333-33 is made up of a core, two 
symmetric jets, and two extended radio lobes \cite{kil1}.
The core
has been identified with the E1 galaxy IC 4296, which has
a redshift $z$=0.013
\cite{slee}. The distance to the source is 35.2 $h^{-1}$ Mpc.
The large-scale structure of PKS 1333-33 has been
studied in detail at 1.3, 2, 6, and 20 cm with an 
angular resolution of 3.2'' \cite{kil1}. The jets are slightly bent, 
presumably as a consequence
of the motion of the core with respect to the intergalactic medium.
The total flux density of the source at 20 cm is $\sim$ 14.5 Jy. The integrated
radio luminosity is $\sim 2.5 \times 10^{41}$ erg s$^{-1}$
assuming a spectral index of $\alpha = 0.6$ and frequency cut-offs at
$10^7$ and $10^{11}$ Hz. Actually, the value $\alpha = 0.6$ is correct
only for the extended radio lobes. The spectral index steepens in the 
jets, reaching $\alpha =1$, while the core has flatter values: 
$\alpha \sim 0.3$.

An interesting feature of PKS 1333-33 is an intense region of synchrotron
emission localized at the outer edge of the eastern lobe. Again, this 
region can be considered as a ``working surface'' formed by the 
deceleration of the jet. This interpretation is supported by VLA 
polarimetric observations, which show a change in the field orientation
from parallel to perpendicular to the jet axis  \cite{kil1}.
This change is probably due to the rearrangement of the field lines in
the post-shock region. The degree of linear polarization in the eastern lobe
is in the range 20\%-40\%. The synchrotron parameters
estimated for this region are: 
the minimum energy density ($\epsilon_{\rm min} = 2.1\times 10^{-13}$ 
erg cm$^{-3}$), the  
minimum magnetic field ($B_{\rm min} = 1.5\times 10^{-6}$ G),   
the minimum total pressure ($P_{\rm min} = 1.1\times 10^{-13}$
dyne cm$^{-2}$), and the 
degree of linear polarization ($m$ = 20-40\%).

The leptonic component of the CRs will
produce synchrotron emission with a spectrum given by 
$S_\nu\propto \nu^{-\alpha}$, where $\alpha = (\gamma -1)/2$.
Since $\alpha = 0.6$, we get $\gamma = 2.2$. 
We assume that electrons and protons in the source obey the same
power law energy spectrum $\propto E^{-\gamma}$.

The degree of linear polarization expected for the synchrotron radiation
when the magnetic field is homogeneous is
\begin{equation}
m_0(\gamma)={3\gamma +3\over{3\gamma +7}}\approx 71\%
\label{homo}
\end{equation}
However, the observed degree of polarization has a mean value of $m(\gamma)\sim
30\%$. This fact can be explained by the presence of a turbulent
component $B_{\rm r}$ in the field, in such a way that
\begin{equation}
m(\gamma)=m_0(\gamma)\;{B_0^2\over{B_0^2+B_{\rm r}^2}}
\label{ran}
\end{equation}
where $B_0$ stands for the homogeneous field. From Eq. (\ref{homo}) and
(\ref{ran}) we get
\begin{equation}
B_{\rm r}\approx 1.2\;B_0
\end{equation}
and consequently $u=B^2_{\rm r}/B^2_{\rm total}\approx 0.6$.

The radius of the acceleration region can be directly measured by means of 
a Gaussian fitting from the
detailed VLA maps obtained by Killeen et al. \cite{kil1}, resulting $R_{\rm hs}
\approx 2.5\;h^{-1}$ kpc. The velocity of the jet is not well established. 
If the source is
$\sim 10^7$ yr old, a velocity $\beta_{\rm jet}\sim 0.2$ can be estimated from
an analysis of the energy budget \cite{kil2}.
As in the case of Cen A,  $a \approx 10^{-2}$.

Taking the above considerations
into account (with a typical value for the total 
magnetic field in the spot of $B_{-5}\sim 10$ \cite{radioastronomie}), we 
obtain from Eq. (\ref{phdsemaxim}) the maximun injection energy for protons, 
$E_{{\rm max}} \approx 6 \times 10^{20}$ eV \cite{pks}. Thus, we have the
following proton injection spectrum
\begin{equation}
Q(E) \propto E^{-2.2},\;\;\;\,E_{0}< E <6 \times 10^{20}\;{\rm eV}.
\label{pacota}
\end{equation}

Using the formalism presented in Sec. 2.2.1 it is straightforward to
compute the main
characteristic of the evolved spectrum. In Fig. 11 we have plotted the 
relation obtained for $\eta$ in the case of
PKS 1333-33. The energy loss by photomeson production creates the
expected cut-off, and the resulting ultra high energy CRs (protons and
neutrons) pile up just
below the threshold energy of photopion production, forming a bump.
However, the energies at which the cut-off and the bump appear seem to
indicate that CRs above 100 EeV might be expected from the location in the 
sky of PKS 1333-33 ($l \approx 313.7^\circ$, $b \approx 27.7^\circ$).

Whether this scenario is the correct one or not should be
answered in a few years by the Pierre Auger Southern Observatory \cite{24,25}
(fluorescence detector plus ground array) as well as by the future eyes of
the OWL \cite{owl} that will deeply watch into the CR-sky.

\section{Hadronic interactions in the final frontier of energy}

The energy spectrum beyond $10^{15}$ eV
needs to be studied indirectly through the extensive air
showers (EAS) CRs produce deep in the atmosphere. The
interpretation of the observed cascades relies strongly on the model
of the shower development used to simulate the transport of particles
through the atmosphere. The parts of the shower model related to
electromagnetic or weak interactions can be calculated with good
accuracy. The hadronic interaction, however, is still subject to
large uncertainties. It generally depends on Monte Carlo
simulations which extrapolate phenomenological models to energies
well beyond those explored at acellerators.

There is a couple of quite elaborate models (the dual parton model
(DPM) \cite{dpm}
and the quark gluon string (QGS) model \cite{qgs})
that provide a complete
phenomenological
description of all facets of soft hadronic processes. These models, inspired
on $1/N$ expansion of QCD are also supplemented with
generally
accepted theoretical principles like duality, unitarity, Regge behavior and
parton structure (for technical details see \cite{werner}). At higher 
energies, however, there is evidence of minijet
production \cite{UA1} and correlation between
multiplicity per event and transverse momentum per particle \cite{ua1},
suggesting that
semihard QCD processes become important in high energy hadronic
interactions. It is precisely the problem of a proper accounting for semihard
processes the major source of uncertainty of extensive air showers event
generators.

Two codes of hadronic interactions with similar
underlying physical assumptions and algorithms tailored for
efficient operation to the highest cosmic ray energies are
SIBYLL \cite{SIBYLL} and QGSJET \cite{QGSJET}.
In these codes, the low $p_{\rm T}$ interactions are modeling by the
exchange of
Pomerons (a hypothetical particle with well defined properties, whose
precise nature in terms of quarks and gluons is not yet completely
understood). Regge singularities are used to determine the momentum
distribution functions of the various sets of constituents, valence
and sea quarks. In the interaction, the hadrons exchange very soft
gluons, simulated by the production of a single pair QCD strings and
the subsequent fragmentation into colour neutral hadrons. In QGSJET 
these events
also involve exchange of multiple pairs of soft strings.

As  mentioned above, the production of small jets is expected to
dominate interactions in the c.m. energy above $\sqrt{s}\approx$ 40 TeV.
The underlying idea behind SIBYLL is that the increase in
the cross section is driven by the production of
minijets \cite{minijet}. The probability distribution for obtaining $N$
jet pairs in a collision  at energy $\sqrt{s}$ is computed regarding elastic $pp$ or
$p\bar{p}$ scattering as a difractive shadow scattering associated
with inelastic processes \cite{durandpi}. The algorithms are
tuned to reproduce the central and fragmentation regions data
up to $p\bar{p}$ collider energies,
and
with no further adjustments they are extrapolated several orders of magnitude.

In QGSJET the theory is formulated entirely in terms of Pomeron
exchanges. The basic idea is to replace the soft Pomeron by a
so-called
``semihard Pomeron'', which is defined to be an ordinary soft Pomeron
with the middle piece replaced by a QCD parton ladder.
Thus, minijets will emerge as a part of the ``semihard Pomeron'',
which is itself the controlling mechanism for the whole interaction.
After
performing the energy sharing among the soft and semihard Pomerons, and
also the sharing
among the soft and hard pieces of the last one; the number of charged
particles
in the partonic cascade is easily obtained generalizing the method of
multiple production of hadrons
as discussed in the QGS model (soft Pomeron showers) \cite{qgs}.

Both, SIBYLL and QGSJET describe particle production in
hadron-nucleus
collisions in a quite similar fashion. The high energy projectile undergoes
a multiple scattering as formulated in Glauber's approach \cite{glauber},
particle
production comes again after the fragmentation of colorless parton-parton
chains
constructed from the quark content of the interacting hadrons.
In cases with more than one wounded nucleon in the target, the extra strings
are connected with sea-quarks in the projectile.
This
ensures that the inelasticity in hadron-nucleus collisions is not much
larger than that corresponding to hadron-hadron collisions.
A higher inelastic nuclear stopping power
yields relatively rapid shower developments which are
ruled out by $p$-nucleus data \cite{frichter}.

\begin{figure}
\begin{center}
\epsfig{file=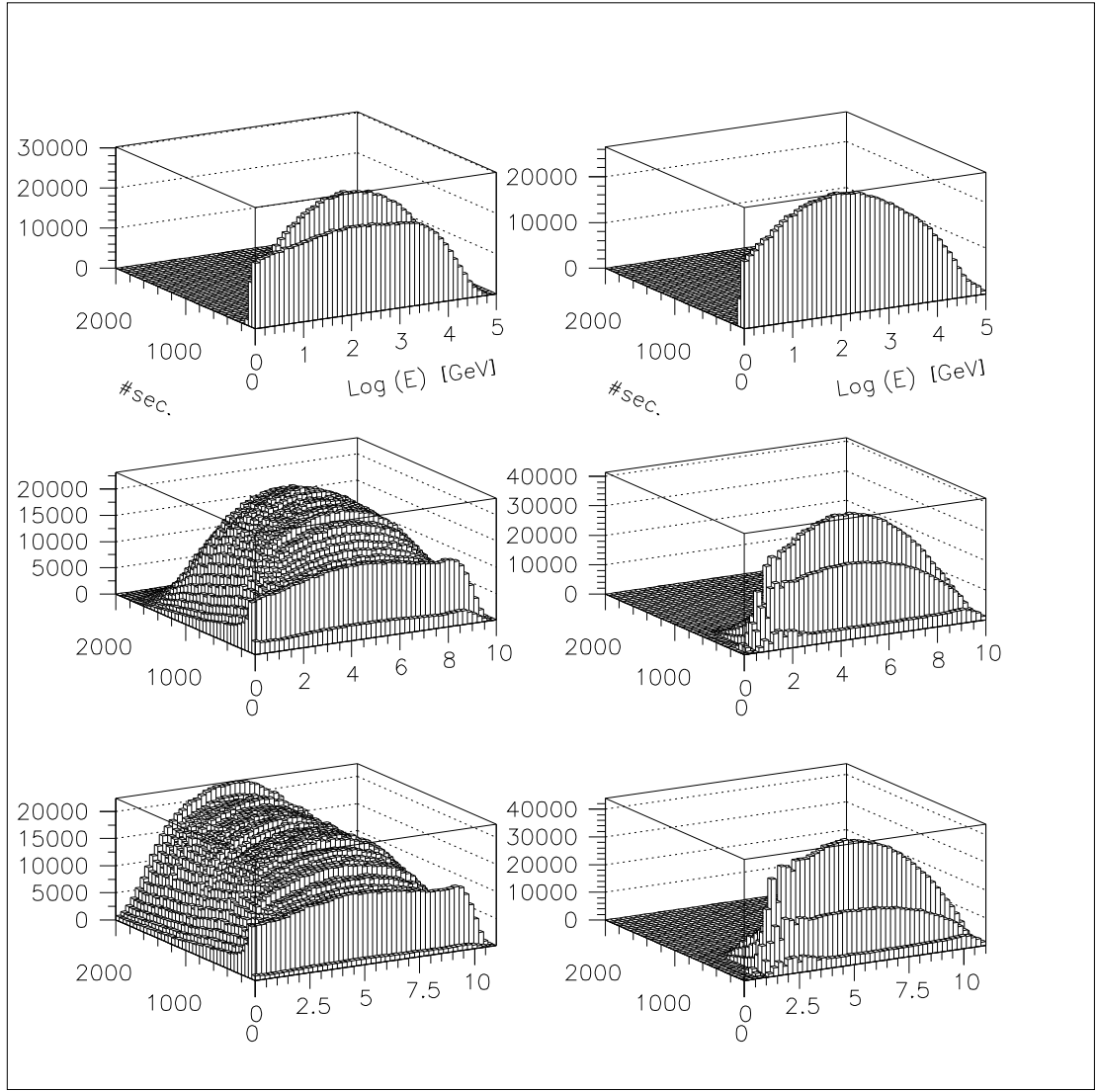,width=15cm,clip=}
\caption{Two--dimension distributions ( Log $(E_{_{\pi^\pm}})$ vs.
number of secondaries) obtained
from $p$-$\bar{p}$ scatterings (incident
energies of $\bar{p}$ 100 TeV -- 10 EeV -- 100 EeV downwards).
In the left hand side we present
the results of QGSJET while the right hand side corresponds to
the ones of SIBYLL.}
\end{center}
\end{figure}

The most direct way to analyze the differences between the models is 
to study the characteristic of the secondaries
generated under similar conditions. For each
hadronic code we generate sets of $10^5$ collisions in order
to analyze the secondaries produced by SIBYLL and QGSJET in $\bar{p}p$ and
$\bar{p}A$ ($A$ represents a nucleus target of mass number $A=10$)
at different projectile energies. Short lived final state particles are
forced to decay according with algorithms included in the SIBYLL and QGSJET
packages. We have recorded the total number of secondaries (baryons, mesons,
and gammas), $N$, produced as a result of the interactions. In all the
considered cases we found that the number of secondaries
coming from QGSJET collisions is larger than the ones corresponding
to the SIBYLL case. For each particle type, two-dimension $N \times$
log$(E)$  distributions were generated. Figure 12 shows selected energy
distributions for secondaries produced in $p$-$\bar{p}$
collisions.\footnote{For details the reader is referred to \cite{prdhi}.}
It is easily seen that when the algorithms are extrapolated several orders
of magnitude, the differences between the predicted number of secondaries 
grow up dramatically. 

Proton induced air showers are generated using AIRES+SIBYLL and
AIRES+QGSJET.\footnote{AIRES is a realistic air shower simulation 
system which includes
electromagnetic interactions algorithms \cite{mocca} and links to the
mentioned SIBYLL and QGSJET models.} Primary energies range 
from $10^{14}$ eV up to $10^{20.5}$
eV. To put into evidence as much as possible the differences between
the intrinsic mechanism of SIBYLL and QGSJET we have always used the
same cross sections for hadronic collisions, namely, the AIRES cross
section.

All hadronic collisions with projectile energies below
200 GeV are
processed with the Hillas Splitting algorithm \cite{hillas}, and the
external
collision package is invoked for all those collisions with energies
above the mentioned threshold. It is worthwhile mentioning that for
ultra-high energy primaries, the
low energy collisions represent a little fraction (no more than 10\% at
 $10^{20.5}$ eV)
of the total number of inelastic hadronic processes that take place
during the shower development.
It is also important to stress that the dependence of the shower
observables on the hadronic model is primarily related to the first
interactions which in all the cases are ultra high energy processes
involving only the external hadronic models.
All shower particles with energies
above the following thresholds were
tracked: 500 keV for gammas, 700 keV for electrons and positrons, 1
MeV for muons, 1.5 MeV for mesons and 80 MeV for nucleons and nuclei.
The particles were injected at the top of the atmosphere (100
km.a.s.l) and the ground level was located at sea level.

\begin{figure}[tb]
\label{xmax}
\begin{center}
\epsfig{file=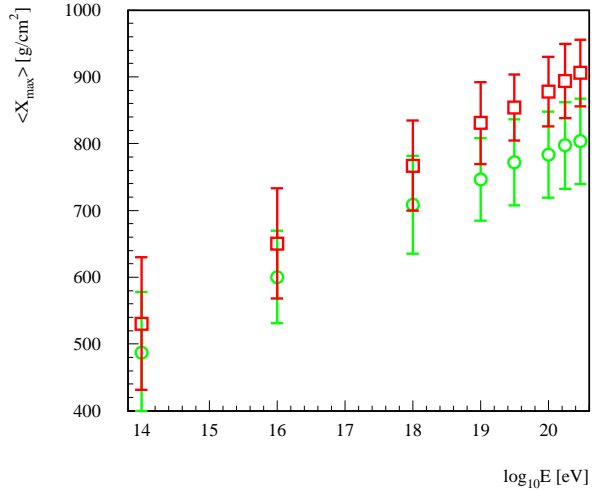,width=8cm,clip=}
\caption{Simulation results for the average slant depth of maximum, $\langle
X_{\rm max}\rangle$, for proton induced showers, plotted versus the
logarithm of the primary energy. The error bars indicate the standard
fluctuations (the RMS fluctuations of the means are always smaller
than the symbols). The squares (circles) correspond to SIBYLL (QGSJET).}
\end{center}
\end{figure}

We have analyzed in detail the longitudinal development of the showers.
The number of different kind of particles have been recorded
as a function of the vertical depth for a number of different observing
levels (more than 100).

The charged multiplicity, essentially electrons and positrons,
is used to determine the number of particles and the location
of the shower maximum by means of four-parameter fits to the
Gaisser-Hillas function \cite{sergio}. In Fig. 13, 
$\langle X_{\rm max}\rangle$ is plotted versus
the logarithm of the primary energy for both, the SIBYLL and QGSJET
cases. It shows up clearly that SIBYLL showers present higher values
for the depth of the maximum, and that the differences between the
SIBYLL and QGSJET cases increase with the primary energy.
This is consistent with the fact that SIBYLL produces less secondaries
than QGSJET --as discussed before-- and as a result,
there is a delay in the electromagnetic shower development which is
strongly correlated with $\pi^0$ decays.
The fluctuations, represented by the error bars, decrease
monotonously as long as the energy increases, passing roughly from 95
g/cm$^2$ at $E=10^{14}$ eV to 70 g/cm$^2$ at $E=10^{20.5}$ eV.

\begin{figure}
\begin{center}
\label{dmu}
\epsfig{file=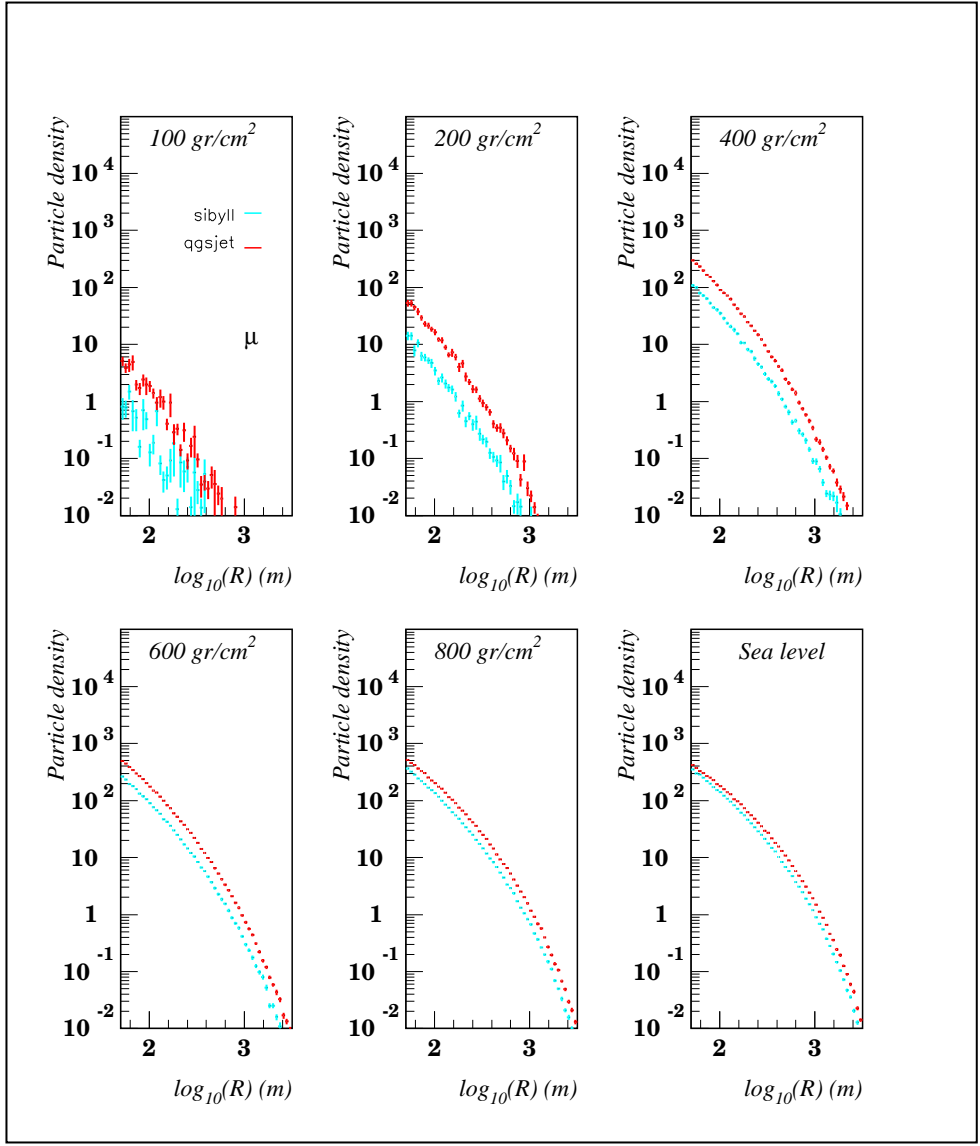,width=13cm,clip=}
\caption{Comparison between the recorded muon
lateral distributions displayed by
SIBYLL  and QGSJET  at different atmospheric altitudes.
Incident primary energy $10^{19}$ eV.}
\end{center}
\end{figure}

Using the recorded particle data,  we have evaluated lateral 
distributions not only at ground altitude but also  at
predetermined observing levels. In the
Figs. 14 and 15 we present the distributions corresponding to a
subset of all
the levels considered, taking into account particles whose distances
to the shower axis are larger than 50 m.

The high-altitude lateral distributions  show
important differences between SIBYLL and QGSJET; such differences
diminish as long as the shower front gets closer to the ground
level. The behaviour can be explained taking into account the
differences between the number of SIBYLL and QGSJET secondaries
previously reported. Due to the fact that SIBYLL produces
less number of secondaries, they have --in average-- more energy and
therefore the number of generations of particles undergoing hadronic
collisions is increased with respect to the QGSJET case. As a result,
during the shower development SIBYLL is called more times than QGSJET,
and this generates a compensation that tends to reduce the difference
in the {\em final\/} number of hadronic secondaries produced during
the entire shower, and consequently in the final decay products, that
is, electrons, gammas and muons.

The lateral distributions of electromagnetic particles are remarkably
similar at both $\langle X_{\rm max}\rangle$
and ground level.\footnote{We want to
stress that $\langle X_{\rm max}\rangle$ is different in each model.} 
However, it comes out from a more detailed analysis of the ground
distributions that they are not strictly coincident and that
the ratio between SIBYLL and QGSJET predictions does depend on $r$,
the distance from the core. In fact, for electrons, this ratio runs
from 1.25 for small $r$ to 0.73 for $r\sim 1000$ m, being equal to 1
at $r \sim 350$ m. A similar behaviour is observed for gammas where
the lateral distributions intersect at $r\sim 1000$ m. In the case of 
lateral muon distributions, QGSJET predicts a higher
density for all distances, but the SIBYLL/QGSJET ratio is not
constant, ranging from 0.74 near the core to 0.56 at 1000 m.

\begin{figure}
\begin{center}
\label{ga}
\epsfig{file=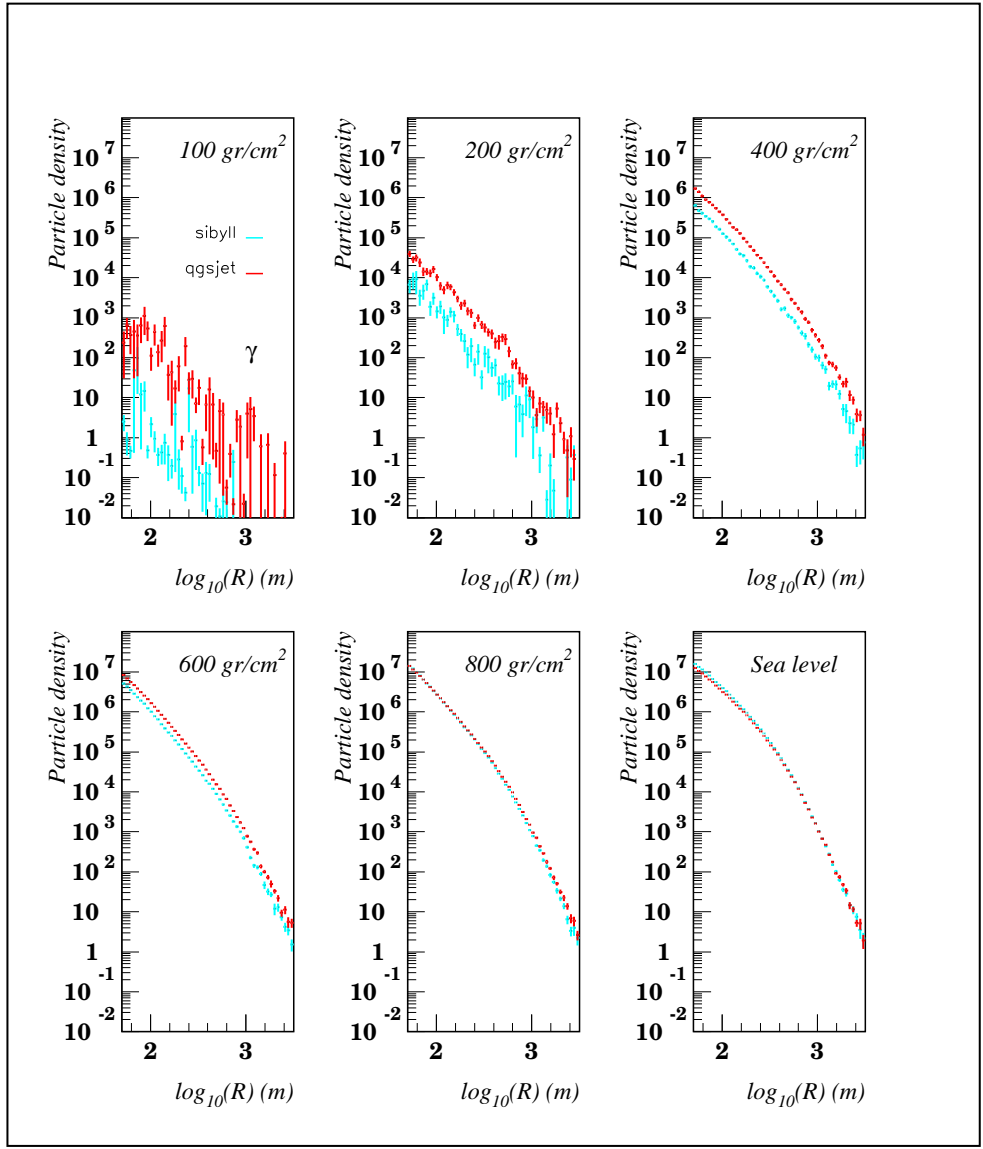,width=13cm,clip=}
\caption{Same as Fig. 14 for the case of gammas.}
\end{center}
\end{figure}

Summing up, the most
oustanding difference between SIBYLL and QGSJET is reflected 
in the predicted number of
secondaries after single $p$-$\bar{p}$ and $\bar{p}$-nuclei
collisions. Such a difference increase steeply with rising
energy. The different
number of secondaries predicted remains noticeable during the first
stages of the shower development, however, the evolution of lateral 
distributions along the longitudinal shower path allows us to clearly
observe how the differences in the distributions become monotonously
damped, yielding rather similar shapes when reaching the ground.
Further, we have shown that the differences observed at ground level
do depend on the distance to the shower core. Consequently, we are
convinced that it will be possible to obtain relevant information
about the hadronic interactions in air showers from the measurement of
particle densities at distances far from (as well as close to) the
shower core. This can be achieved if CR experiments are designed
with appropiate dynamic ranges.

Measurement of particle numbers at high atmospheric altitudes (fluorence
detectors) together with shower maximun and changes rates, should contribute
positively in the understanding of hadronic interactions if the kind of cosmic
particle that induces the shower could be determined separately.
Finally, it should be noted that most of the model discrepancies in hadronic
interactions will be naturally reduced with the help of data
obtained from future accelerator experiments like the well known Large
Hadron Collider (LHC).

\section{Looking Forward}

Cosmic rays of extremely high energy are a weird phenomenon of
nature for which no truly satisfactory explanation has been found yet.  
This may simply reflect our present ignorance of conditions or processes in
some highly energetic regions of the Universe - or may imply that
exotic mechanisms are at play. 

After thirty years of careful work by several groups all over
the world, we are in possession of a tantalizing body of data,
sufficient to raise our curiosity and wonder, but not to succeed
unravelling in the mistery. The coming avalanche of high quality 
CR-observations promises to make the begining of the next millenium  
an extremely exciting period for CR-physics.

\section*{Acknowledgements}

This is a summary of the Thesis presented as partial fulfillment of the
Doctoral degree requirement in the Department of Physics at the
University of La Plata. I would like to express 
my gratitude to M. T.
Dova and L. N. Epele for their guidance throughout this ongoing
project and for invaluable discussions and collaboration. This Thesis
would have been imposible without their effort. 
It is a pleasure to thank my friends, D. A. G\'omez Dumm, and D.
F. Torres  for critical reading of this summary as
well as the whole spanish manuscript. Any errors as well as the
overall style in which this summary is written still belongs (for
good or bad) entirely to me. I am particulary grateful 
to  J. Combi, H. Fanchiotti, C. Garc\'{\i}a Canal, S. Perez Bergliaffa, 
G. Romero, E. Roulet, S. Sciutto, J. Swain, M. Trobo, H. Vucetich and 
F. Zyserman for enlightening discussions. The up-to-date cosmic 
ray data sample was kindly provided by C. Hojvat. The work has been 
partially supported by FOMEC.

\end{document}